%% file: flux_rapids.tex
\documentclass{jfm}
\usepackage{graphicx,xcolor}
\usepackage{amssymb,latexsym}
\usepackage{mathrsfs}
\usepackage{soul}
\usepackage{natbib}
\usepackage{amsbsy}
\usepackage{psfrag}
\usepackage{mathtools}
\usepackage{color}
\usepackage{subfig}
\input{macros.tex}

\shorttitle{Coherent structure of the kinetic energy transfer}
\shortauthor{S. Dong, Y. Huang, X. Yuan, and A. Lozano-Dur\'an}

\title{The coherent structure of the kinetic energy transfer in shear turbulence}

\author{Siwei Dong\aff{1},
        Yongxiang Huang\aff{2},
        Xianxu Yuan\aff{1},
 and Adri\'an Lozano-Dur\'an\aff{3} \corresp{\email{adrianld@stanford.edu}}}

\affiliation{
\aff{1}State Key Lab. of Aerodynamics, China Aerodynamics R\&D Center, 621000 Mianyang, China
\aff{2}State Key Lab. of Marine Environmental Science, Xiamen U., 361102 Xiamen, China
\aff{3}Center for Turbulence Research, Stanford University, Stanford CA, 94305 USA
}

\begin{document}

\maketitle

\begin{abstract}
The cascade of energy in turbulent flows, i.e., the transfer of
kinetic energy from large to small flow scales or vice versa (backward
cascade), is the cornerstone of most theories and models of turbulence
since the 1940s. Yet, understanding the spatial organisation of
kinetic energy transfer remains an outstanding challenge in fluid
mechanics.  Here, we unveil the three-dimensional structure of the
energy cascade across the shear-dominated scales using numerical data
of homogeneous shear turbulence.  We show that the characteristic flow
structure associated with the energy transfer is a vortex shaped as an
inverted hairpin followed by an upright hairpin. The asymmetry between
the forward and backward cascade arises from the opposite flow
circulation within the hairpins, which triggers reversed patterns in
the flow.
\end{abstract}


\section{Introduction}\label{sec:intro}

Turbulence exhibits a wide range of flow scales, whose interactions
are far from understood \citep{Cardesa2017}.  These interactions are
responsible for the cascading of kinetic energy from large eddies to
the smallest eddies, where the energy is finally dissipated
\citep{Richardson1922, Obukhov1941, Kolmogorov1941, Kolmogorov1962,
  aoyama:ishihara:2005, Falkovich2009}. Given the ubiquity of
turbulence, a deeper understanding of the energy transfer among the
flow scales would enable significant progress to be made across
various fields ranging from combustion \citep{Veynante2002},
meteorology \citep{Bodenschatz2015}, and astrophysics
\citep{Young2017} to engineering applications of external aerodynamics
and hydrodynamics \citep{Sirovich1997, Hof2010, Marusic2010,
  Kuhnen2018, Ballouz2018}.  In the vast majority of real-world
scenarios, turbulence is accompanied by an abrupt increase of the mean
shear in the vicinity of the walls due the friction induced by the
latter. These friction losses are responsible for roughly 10\% of the
electric energy consumption worldwide \citep{Kuhnen2018}.  Moreover,
the success of large-eddy simulation (LES), which is an indispensable
tool for scientific and engineering applications \citep{Bose2018},
lies in its ability to correctly reproduce energy transfer among
scales.  Hence, a comprehensive analysis of the interscale energy
transfer mechanism is indispensable for both physical understanding of
turbulence and for conducting high-fidelity LES.

Substantial efforts have been directed toward the statistical
characterisation of interscale kinetic energy transfer using flow data
acquired either by simulations or experimental measurements
\citep[e.g.,][]{natrajan:christensen:2006:pof,Kawata2018}.  Several
works have further examined the cascading process conditioned to
selected regions of the flow, mainly motivated by the fact that the
interscale energy transfer is highly intermittent both in space and
time \citep[see][and references therein]{pio:cab:moin:lee:1991,
  Domaradzki1993, Cerutti1998, aoyama:ishihara:2005, Ishihara2009,
  dubrulle2019}. By conditionally averaging the flow, previous works
have revealed that kinetic energy fluxes entail the presence of shear
layers, hairpin vortices, and fluid ejections/sweeps.  However,
further progress in the field has been hindered by the scarcity of
flow information, which has been limited to a few velocity components
and two spatial dimensions. Consequently, less is known about the
underlying three-dimensional structure of the energy transfer, which
is the focus of this work.

\citet{har:kle:ung:fri:1994:pof} conducted one of the earliest
numerical investigations on kinetic energy fluxes and their
accompanying coherent flow structures. Their findings showed that the
backward transfer of energy is confined within a near-wall shear
layer.  In a similar study, \citet{pio:yu:adrian:1996} proposed a
model comprising regions of strong forward and backward energy
transfer paired in the spanwise direction, with a quasi-streamwise
vortex in between. This view was further supported by an LES study on
the convective planetary boundary layer \citep{lin:1999:pof}.  The
previous works pertain to the energy transfer across flow scales under
the influence of the mean shear, which is the most relevant case from
the engineering and geophysical viewpoints.  Still, it is worth
mentioning that the inertial energy cascade has been classically
ascribed to the stretching exerted among vortices at different scales
in isotropic turbulence \citep{Goto2017, Motoori2019}, although recent
works have debated this view in favour of strain-rate
self-amplification as the main contributor to the energy transfer
among scales \citep{car:bra:2019}. The reader is referred to
\citet{Alexakis2018} for an unified and exhaustive review of the
different energy transfer mechanisms.

On the experimental side, Port\'e-Agel and collaborators
\citep{porte:par:mene:eich:2001, porte:pahlow:meneveau:par:2001,
  carper:portel:2004:jot} performed a series of studies in the
atmospheric boundary layer.  They conjectured that regions of intense
forward and backward cascades organised around the upper trailing edge
and lower leading edge of a hairpin, respectively. Later
investigations using particle-image velocimetry in turbulent boundary
layers with smooth walls \citep{natrajan:christensen:2006:pof} and
rough walls \citep{hong:katz:meneveau:schultz:2012:jfm} corroborated
the presence of counter-rotating vertical vorticity around regions of
intense kinetic energy transfer, consistent with
\citet{carper:portel:2004:jot}. More recent studies of the mixing
layer induced by Richtmyer--Meshkov instability
\citep{liu:xiao:2016:pre} have also revealed flow patterns similar to
those described above.

Previous numerical and experimental studies have helped advance our
understanding of the spatial structure of energy transfer; however,
they are limited to only two dimensions.  With the advent of the
latest simulations and novel flow identification techniques
\citep[e.g.][]{del:jim:zan:mos:06, loz:flo:jim:12, Lozano2014,
  don:lon:seki:jim:2017:jfm, osawa:jimenez:2018}, the
three-dimensional characterisation of turbulent structures is now
achievable to complete the picture.  In the present study, we shed
light on the three-dimensional flow structure associated with regions
of intense energy fluxes in the most fundamental set-up for shear
turbulence.

\textcolor{black}{The present work is organised as follows. The
  numerical database and the filtering approach to study the energy
  transfer in homogeneous shear turbulence is described in
  \S\ref{sec:method}. The results are presented in \S\ref{sec:stat},
  which is further subdivided into two parts. In
  \S\ref{sec:rela_posi}, we show the spatial organisation of the flow
  structures responsible for the forward and backward energy
  cascade. The coherent flow associated with both energy cascades is
  analysed in \S\ref{sec:coherent}. Finally, conclusions are offered
  in \S\ref{sec:conclude}.}

\section{\textcolor{black}{Numerical experiment and filtering approach}}
\label{sec:method}

\subsection{\textcolor{black}{Database of homogeneous shear turbulence}}

\begin{figure}
\vspace{1cm}
	\centerline{
        \includegraphics[width=0.6\textwidth]{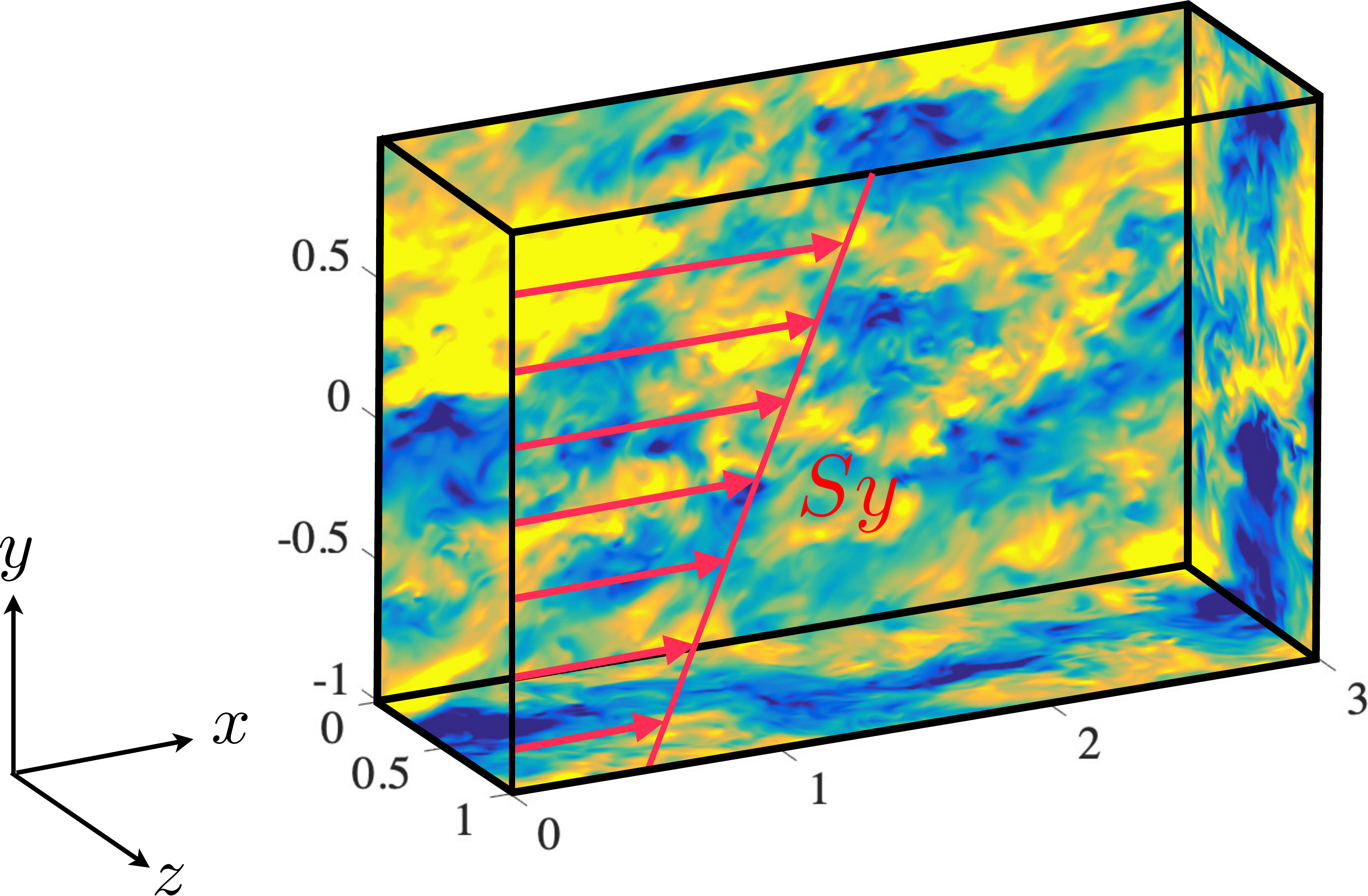}%
	}
	\caption{Schematic of the computational domain.  $S$ is the
          superimposed mean shear.  The read arrows are the streamwise
          mean velocity profile, $U=Sy$. The colours represent the
          magnitude of streamwise fluctuating velocity, $u$, in the
          range $[-1,1]$ from yellow to blue.  The streamwise,
          vertical, and spanwise coordinates are $x$, $y$, and $z$,
          respectively.  \textcolor{black}{The lengths are normalised
            with the spanwise size of the domain}. The velocities are
          normalised with the friction velocity defined as
          $\utau^2=|\nu S-\bra uv \ket|$. The flow is periodic in $x$
          and $z$.  Periodicity is enforced at the upper and lower
          boundaries for points that are uniformly shifted in the $x$
          consistently with the superimposed mean shear.}
	\label{fig:setup}
\end{figure}

We examine data from the direct numerical simulation (DNS) of
statistically stationary, homogeneous, shear turbulence (SS-HST) from
\cite{sek:don:jim:15}. The flow is defined by turbulence in a doubly
periodic domain with a superimposed linear mean shear profile
\citep{Champagne1970}. This configuration, illustrated in figure
\ref{fig:setup}, is considered the simplest anisotropic flow, sharing
the natural energy-injection mechanism of real-world shear
flows. Hence, our numerical results are utilised as a proxy to gain
insight into the physics of wall-bounded turbulence without the
complications of the walls \citep{don:lon:seki:jim:2017:jfm}.
\textcolor{black}{The Reynolds number of the simulation based on the
  Taylor microscale is $Re_{\lambda} =
  q^2(5/3\nu\varepsilon)^{1/2}\approx 100$ (with $q^2$ and
  $\varepsilon$ the kinetic energy and dissipation, respectively),
  which is comparable to that in the logarithmic layer of wall-bounded
  turbulence at a friction Reynolds number of $Re_{\tau}\approx
  2000$.}

Hereafter, fluctuating velocities are denoted by $u$, $v$, and $w$ in
the streamwise ($x$), vertical ($y$), and spanwise ($z$) directions,
respectively.  The mean velocity vector averaged over the homogeneous
directions and time is $(U,V,W)=(Sy,0,0)$, where $S$ is the constant
mean shear rate.  Occasionally, we use subscripts $1$, $2$, and $3$ to
refer to the streamwise, vertical, and spanwise directions (or
velocities), respectively, in which case repeated indices imply
summation.  \textcolor{black}{Details of the simulation are listed in
  table \ref{table:info:dns}.}
\begin{table} \begin{center} \begin{tabular}{lcccccccccc}
\textcolor{black}{$Re_{\lambda}$} & \textcolor{black}{$ A_{xz} $} & \textcolor{black}{$A_{yz}$} &
\textcolor{black}{$\Delta x/\eta$}  & \textcolor{black}{$\Delta y/\eta$} & \textcolor{black}{$\Delta z/\eta$} & \textcolor{black}{$L_c/\eta$} & \textcolor{black}{$L_z/\eta$}
\\[1ex]
104  & 3 & 2 & 1.6$\pm$0.18 & 1.0$\pm$0.12 & 1.0$\pm$0.12& 36 & 366 \\ [1ex]
\end{tabular}
\end{center}
\caption{\textcolor{black}{Simulation parameters for the DNS of SS-HST
    used in this paper. $Re_{\lambda}=q^2(5/3\nu\varepsilon)^{1/2}$ is
    the Reynolds number based on the Taylor microscale ($q^2$ and
    $\varepsilon$ are the kinetic energy and dissipation,
    respectively); $A_{xz}=L_x/L_z$ and $A_{yz}=L_y/L_z$ are the box
    aspect ratios of the computational domain; $\Delta x/\eta$,
    $\Delta y/\eta$ and $\Delta z/\eta$ are the grid resolutions in
    terms of the average Kolmogorov length-scale $\eta$, with their
    standard deviations due to intermittency. $\Delta x$ and $\Delta
    z$ are computed from the number of Fourier modes before
    dealiasing; $L_c$ is the Corrsin length-scale defined as $L_c =
    \sqrt{\varepsilon/S^3}$.}}
\label{table:info:dns}
\end{table}

\textcolor{black}{The code integrates in time the equation for the
  vertical vorticity $\omega_y$ and for the Laplacian of $v$.  The
  spatial discretization is dealiased Fourier
  spectral in the two periodic directions, and compact finite
  differences with spectral-like resolution in $y$. The Navier--Stokes
  equations of motion, including continuity, are reduced to the evolution
  equations for $\phi =\nabla^2 v$ and $\omega_y$
  \citep{KimMoinMoser1987} with the advection by the mean flow
  explicitly separated,
\beq
\pdif{\omega_y}{t} + Sy\pdif{\omega_y}{x} = h_g + \nu \nabla^2 \omega_y,
\qquad
\pdif{\phi }{t} + Sy\pdif{\phi }{x} = h_v + \nu \nabla^2 \phi,
\label{eq:vorphi}
\eeq
where $\nu$ is the kinematic viscosity. Defining $\Hvec = \uvec \times \omgvec$,
\beq
h_g \equiv  \pdif{H_x}{z} - \pdif{H_z}{x} - S\pdif{v}{z},
\qquad
h_v \equiv - \pdif{}{y} \left( \pdif{H_x}{x} + \pdif{H_z}{z} \right)
            + \left( \pddif{}{x} + \pddif{}{z} \right) H_y.
\la{eq:hgv}
\eeq
In addition, the governing equation for $\langle \uvec \rangle_{xz} $ are
\beq
\pdif{\langle u \rangle_{xz}}{t} =
-\pdif{\langle u v\rangle_{xz}}{y} + \nu \pddif{\langle u \rangle_{xz}}{y},
\qquad
\pdif{\langle w \rangle_{xz}}{t} =
-\pdif{\langle w v\rangle_{xz}}{y} + \nu \pddif{\langle w \rangle_{xz}}{y},
\label{eq:u00w00}
\eeq 
where $\langle \cdot \rangle_{xz}$ denotes averaging on the
homogeneous directions.  The time stepping is a third-order explicit
Runge--Kutta \citep{Spalart1991} modified by an integrating factor for
the mean-flow advection.}

\textcolor{black}{The spanwise, vertical, and spanwise size of the
  domain are denoted by $L_x$, $L_y$, and $L_z$, respectively. The
  numerical domain is periodic in $x$ and $z$, with boundary
  conditions in $y$ that enforce periodicity between uniformly
  shifting points at the upper and lower boundaries. More precisely,
  the boundary condition used is that the velocity is periodic between
  pairs of points in the top and bottom boundaries of the
  computational box, which are shifted in time by the mean shear
  \citep{Baron1982, Schumann1985, GerzSchumannElghobashi1988,
    BalbusHawley1998}. For a generic fluctuating quantity $g$,
\beq
g(t, x, y, z) = g(t, x + m S t  L_y + l L_x, y + m L_y , z + n L_z),
\la{eq:bc1}
\eeq
where $l,m$ and $n$ are integers.  In terms of the spectral
coefficients of the expansion,
\beq
g(t, x,y,z) = \sum_{k_x} \sum_{k_z} \widehat{g}(t,k_x,y,k_z) \exp[\ii (k_x x + k_z z)],
\label{eq:shear-periodic-fou}
\eeq
the boundary condition becomes
\beq
\widehat{g}(t, k_x, y, k_z) = \widehat{g}(t, k_x, y + m L_y, k_z) \exp[\ii k_x m S t L_y],
\label{eq:sp_bc}
\eeq
where $k_i = n_i \Delta k_i\, (i=x,z)$ are wavenumbers, $n_i$ are
integers, and $\Delta k_i = 2\pi/L_i$. This shifting boundary
condition in $y$ avoids the periodic remeshing required by
tilting-grid codes \citep{Rogallo81}, and most of their associated
enstrophy loss.}

The simulations are characterised by the streamwise and vertical
aspect ratios of the simulation domain, $A_{xz}=L_x/L_z=3$ and
$A_{yz}=L_y/L_z=2$, and the Reynolds number $Re_z=SL_z^2/\nu$, where
$\nu$ is the kinematic viscosity of the fluid. The velocities are
normalised with the friction velocity defined as $\utau^2=|\nu S-\bra
uv \ket|$. Occasionally, we also use the Corrsin
  length, $L_c=(\varepsilon/S^3)^{1/2}$, above which the mean shear
  dominates, where $\varepsilon$ is the mean rate of turbulence
  kinetic energy dissipation.

\subsection{\textcolor{black}{Definition of interscale kinetic energy transfer $\varPi$}}

\textcolor{black}{The evolution equation for the $i$-th component of the velocity is
\begin{equation}\label{eq:u_i}
\frac{\partial u_i}{\partial t}+\frac{\partial u_iu_j}{\partial x_j}+Sy\frac{\partial u_i}{\partial x_1}=-Sv\delta_{i1}-\frac{1}{\rho}\frac{\partial p}{\partial x_i}+\nu\frac{\partial^2 u_i}{\partial x_j \partial x_j}.
\end{equation}
After low-pass filtering (\ref{eq:u_i}) using an isotropic Gaussian
filter with filter size $r_f$, the equation for the filtered kinetic
energy $\widetilde{k}=\widetilde{u}_i\widetilde{u}_i/2$ is given by
\begin{equation}
\frac{D \widetilde{k}}{D t}=-S\widetilde{u}_1\widetilde{u}_2-J-\nu\frac{\partial \widetilde{u}_i}{\partial x_i}\frac{\partial \widetilde{u}_i}{\partial x_i}-\varPi_{\mathrm{MF}}-\varPi,
\end{equation}
where $D(\cdot)/Dt$ is the material derivative and 
\begin{eqnarray}
\varPi_{\mathrm{MF}}&=&-[\widetilde{u_iU}-\widetilde{u}_i\widetilde{U}]\frac{\partial \widetilde{u}_i}{\partial x_1}, \\
\varPi&=&-\tau_{ij}\frac{\partial \widetilde{u}_i}{\partial x_j}=-[\widetilde{u_iu_j}-\widetilde{u}_i\widetilde{u}_j]\frac{\partial \widetilde{u}_i}{\partial x_j},\\
J &=& \frac{\partial \widetilde{u}_j\widetilde{u}_i^2/2}{\partial x_j} 
+\frac{\partial \widetilde{Sy}\widetilde{u}_i^2/2}{\partial x}
+\frac{1}{\rho}\frac{\partial \widetilde{p}\widetilde{u}_i}{\partial x_i}
-\nu\frac{\partial }{\partial x_j}(\widetilde{u}_i\frac{\partial \widetilde{u}_i}{\partial x_j})+\\
&+&\frac{\partial \widetilde{u}_i\widetilde{u_iu_j} }{\partial x_j}
+\frac{\partial \widetilde{u}_i^2\widetilde{u}_j}{\partial x_j} 
+\frac{\partial \widetilde{u}_i\widetilde{Syu_i} }{\partial x}-
\frac{\partial \widetilde{Sy}\widetilde{u}_i^2}{\partial x}. \nonumber
\end{eqnarray}
$\varPi_{\mathrm{MF}}$ is the kinetic energy transfer due to the
interaction between the filtered and mean flow, $\varPi$ is the
transfer between flow scales, and $J$ is a spatial flux. We are
concerned with $\varPi$, which represents the energy
cascade. A positive value of $\varPi$ implies a transfer of energy
from the large, unfiltered scales to the small, filtered scales
(forward cascade), while negative values of $\varPi$ represent an
opposite transfer (backward cascade). Filtered quantities are
calculated using filter widths $r_f$ ranging from $23.3\eta$ in the
viscous range to $80.0\eta$ in the inertial range, where $\eta$ is the
Kolmogorov length scale. Since the large-scale structures in SS-HST
are only slightly elongated in the streamwise direction
\citep{don:lon:seki:jim:2017:jfm}, we utilise an spatially isotropic
filter.}
\begin{figure}
\vspace{1cm}
	\centerline{
        \includegraphics[width=1\textwidth]{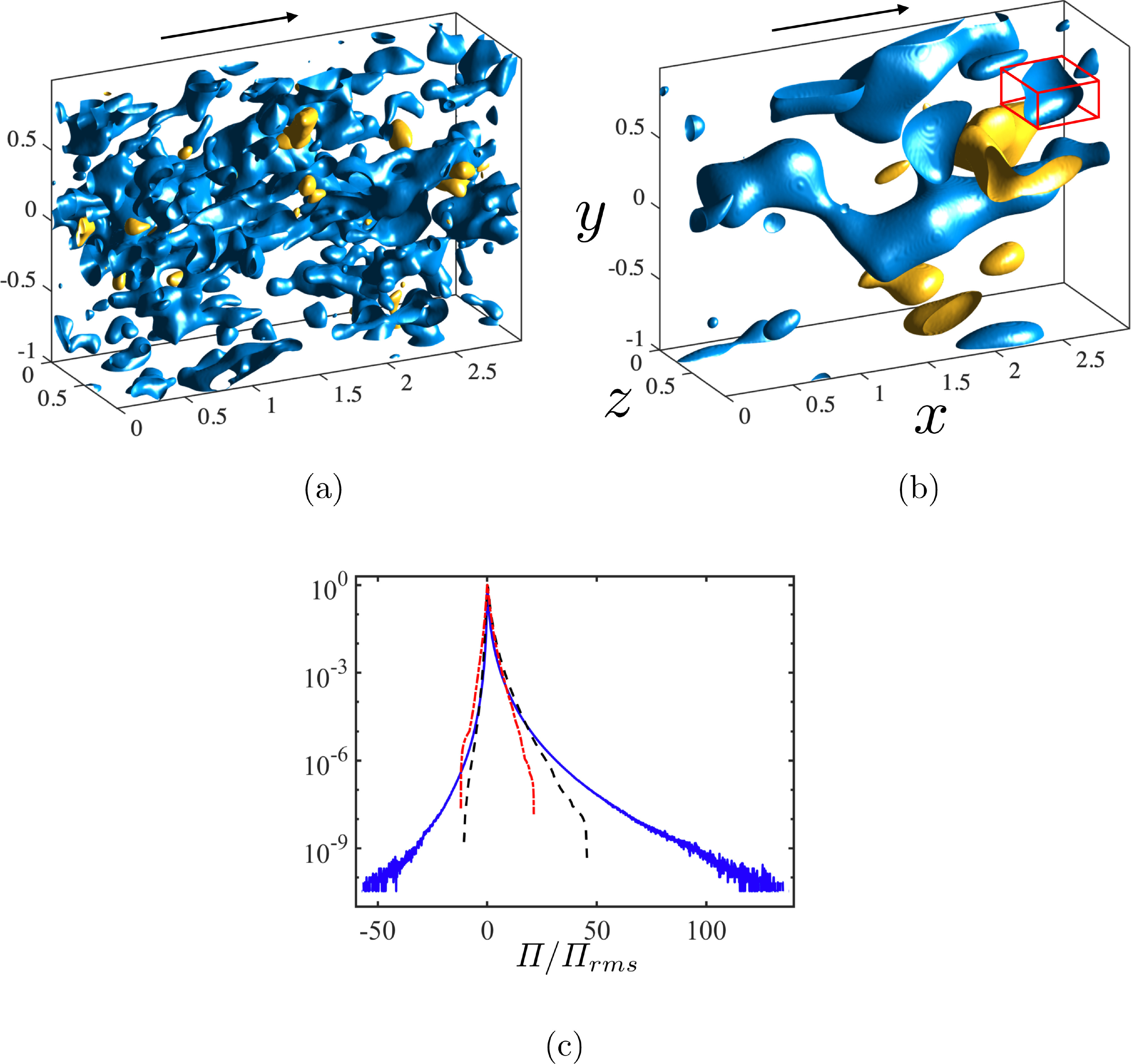}
	}
	\caption{ Instantaneous spatial distribution of regions of
          intense positive (blue) and negative (yellow) kinetic energy
          transfer $\varPi$ for filter sizes equal to (a)
          $r_f=23.3\eta$, and (b) $r_f=60\eta$. The arrow indicates
          the mean flow direction. \textcolor{black}{The lengths are
            normalised with the spanwise size of the domain}.  The red
          box in panel (b) is the bounding box of an individual
          $\varPi^+$-structure identified by Eq. (\ref{eq:def:flux}).
          (c) The p.d.f. of $\varPi$ for filter sizes equal to
          $23.3\eta$ ($\textcolor{black}{\dashed}$) and $60\eta$
          ($\chndot$).  For
            comparison purposes, the figure also includes the
            additional filter width $r_f=5.8\eta$}
          ($\textcolor{black}{\solid}$).  Each p.d.f. is normalised by
          its standard deviation, $\varPi_{rms}$.  
	\label{fig:flux}
\end{figure}

\section{Results}\label{sec:stat}

Figures \ref{fig:flux}(a) and \ref{fig:flux}(b) show the instantaneous
spatial distribution of regions of intense $\varPi$ for two filter
widths, $r_f=23.3\eta$ and $r_f=60\eta$. Both forward and backward
cascades coexist, as seen from the positive and negative regions of
$\varPi$, although the forward cascade prevails. The spatial
distribution of the fluxes is strongly inhomogeneous, with regions of
intense energy transfer organised into intermittent spots
\citep{pio:cab:moin:lee:1991, Domaradzki1993, Cerutti1998,
  aoyama:ishihara:2005, Ishihara2009, Wu2017, Yang2017}. The
probability density function (p.d.f.) of $\varPi$ (figure
\ref{fig:flux}c) shows that the skewness factor of $\varPi$ decreases
monotonically with the filter size (from 6.54 to
0.73), i.e., the cascade process becomes more
symmetric at larger scales. The intensity of rare events also
decreases dramatically as a function of the filter size.

In the following, we study the properties of three-dimensional
structures of intense kinetic energy transfer. Individual structures
are identified as a contiguous region in space satisfying
\begin{equation}\label{eq:def:flux}
|\varPi(\xvec)|>\alpha \varPi_{rms},
\end{equation}
where $\xvec=(x,y,z)$, $\alpha>0$ is a threshold parameter, and
$\varPi_{rms}$ is the standard deviation of $\varPi$. The value of
$\alpha$ is chosen to be 1.0 based on a percolation analysis
\citep{moi:jim:04}; however, similar conclusions are drawn for
$0.5<\alpha<2.0$. Connectivity is defined in terms of the six
orthogonal neighbours in the Cartesian mesh of the DNS. By
construction of Eq. (\ref{eq:def:flux}), each individual structure
belongs to either a region of forward or backward cascade, denoted by
$\varPi^+$ and $\varPi^-$, respectively. \textcolor{black}{Each
  structure is circumscribed within a box aligned to the Cartesian
  axes, whose streamwise, vertical, and spanwise sizes are denoted by
  $\Delta_x$, $\Delta_y$, and $\Delta_z$,
  respectively. The diagonal of the bounding box is given by
  $d=\sqrt{\Delta_x^2 + \Delta_y^2 + \Delta_z^2}$. The total number of
  structures used to compute the averaged flow field is of the order
  of $10^4$.} Examples of instantaneous structures for two different
filter sizes can be seen in figure \ref{fig:flux}(a) and figure
\ref{fig:flux}(b).  In the latter, one individual structure of
$\varPi^+$ is enclosed within its bounding box (in red).

\begin{figure}
\vspace{0.5cm}
\centerline{
  \includegraphics[width=0.98\textwidth]{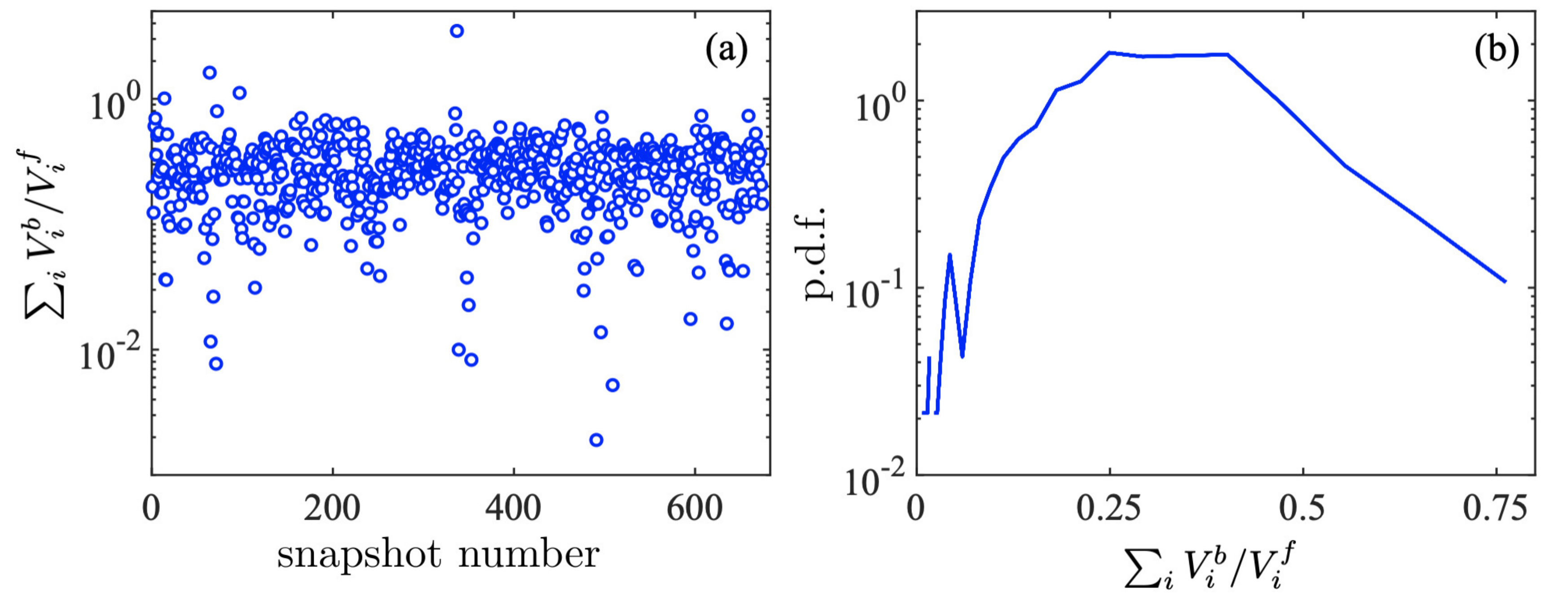}
\vspace{0.1cm}
}
\caption{(a) The ratio of the total volume of
    backward cascade structures $\sum_i V_i^b$ and total volume of
    forward cascade structures $\sum_i V_i^f$ as a function of the
    snapshot used in the analysis. $V_i^f$ and $V_i^b$ are the volume
    of individual forward and backward structures, respectively,
    satisfying (\ref{eq:def:flux}) with $\alpha=1.0$.  The results are
    for filter width $r_f=60\eta$.  (b) The p.d.f. of the volume ratio
    of backward and forward cascade structures from the data shown in
    panel (a).}
\label{fig:ratio_fb}
\end{figure}

\textcolor{black}{Prior to the investigation of the spatial
  organisation of the energy transfer, we discuss the amount of
  forward and backward cascade structures. Figure
  \ref{fig:ratio_fb}(a) shows the ratio of the total volume of
  backward cascade structures ($\sum_i V_i^b$) and
  the total volume of forward cascade structures
  ($\sum_i V_i^f$) for the different times employed in the analysis
  and for $r_f=60\eta$. The p.d.f. of the volume ratio $\sum_i
  V_i^b/\sum_i V_i^f$ is given in figure \ref{fig:ratio_fb}(b). The
  mean value is roughly 0.3, i.e., forward cascade events dominate,
  consistent with previous results in the literature
  \citep{pio:cab:moin:lee:1991, aoyama:ishihara:2005}. However, the
  ratio varies widely among instants, ranging from $2\times 10^{-3}$
  to $4$, showing the time intermittency of the cascade.}

\subsection{Spatial organisation of the energy cascades}\label{sec:rela_posi}

The spatial organisation of $\varPi^+$ and $\varPi^-$ is studied
through the three-dimensional joint p.d.f., $p^{ij}$, of the relative
distances between the individual structures of type $i$ and $j$, where
$i$ and $j$ refer to either $\varPi^+$ or $\varPi^-$. The vector of
relative distances is defined as
\begin{equation}\label{eq:posi}
\deltavec^{(ij)} = (\delta_x,\delta_y,\delta_z)^{(ij)} = 2\frac{\xvec_c^{(j)}-\xvec_c^{(i)}}{d^{(j)}+d^{(i)}},
\end{equation}
where $\xvec_c^{(i)}$ is the centre of gravity of one individual
structure, and $d^{(i)}$ is the diagonal length of its bounding box
(highlighted in red in figure \ref{fig:flux}b). Only pairs of
structures with similar sizes are considered in the computation of
relative distances \citep{loz:flo:jim:12,osawa:jimenez:2018,
  don:lon:seki:jim:2017:jfm}, in particular, those satisfying $1/2 \le
d^{(j)}/d^{(i)} \le 2$. We also take advantage of the spanwise
symmetry of the flow, and $\delta_z$ is chosen to be positive toward
the closest $j$-type structure.
 
\begin{figure}
\vspace{0.5cm}
    \centerline{
        \psfrag{X}{$\delta_z$}\psfrag{Y}{$\delta_y$}%
		\includegraphics[width=0.45\textwidth]{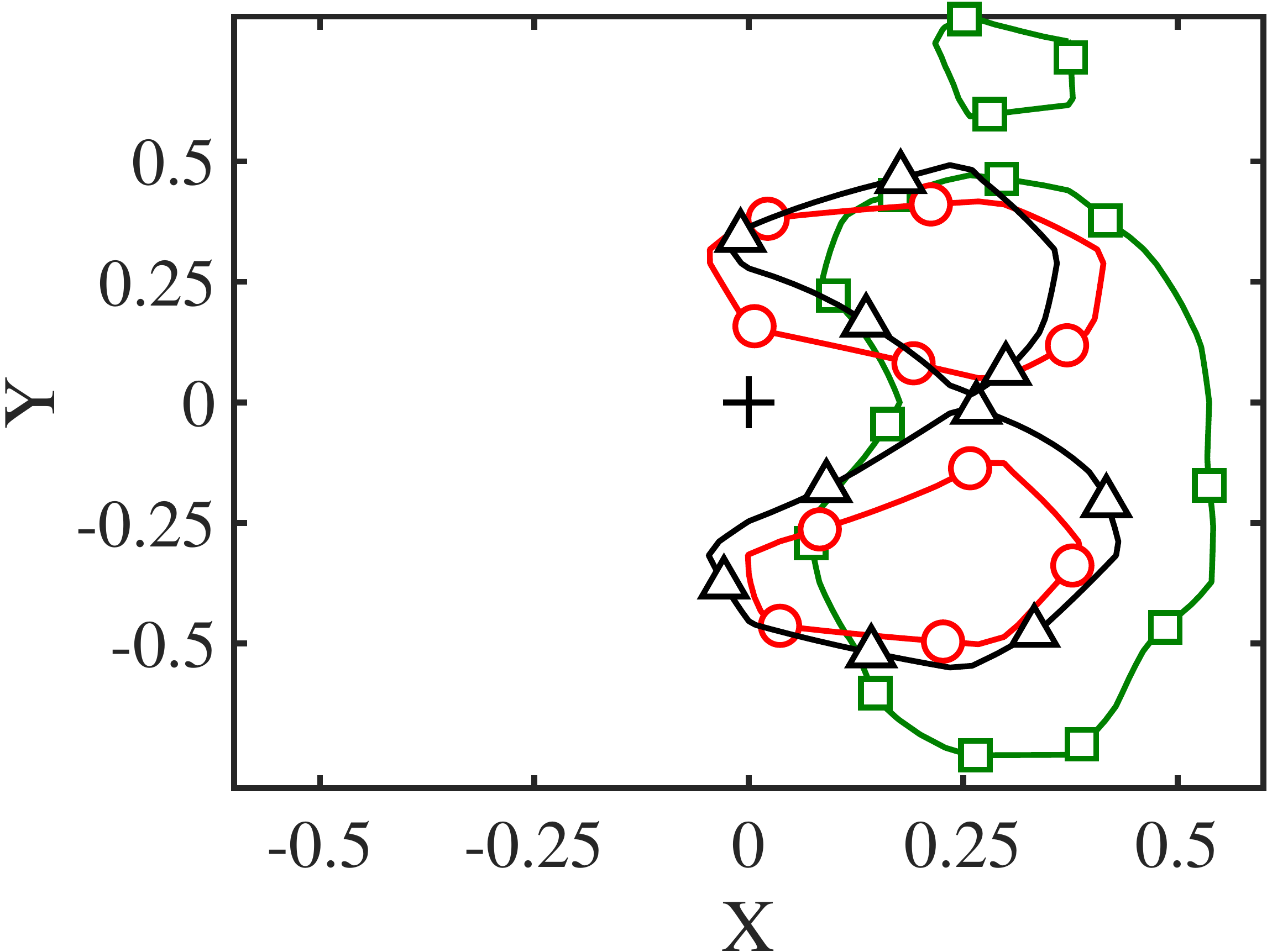}%
		\mylab{-4.5cm}{4cm}{(\aaa)}%
       	        \vspace{0.5cm}
		\psfrag{X}{$\delta_x$}\psfrag{Y}{}%
		\includegraphics[width=0.45\textwidth]{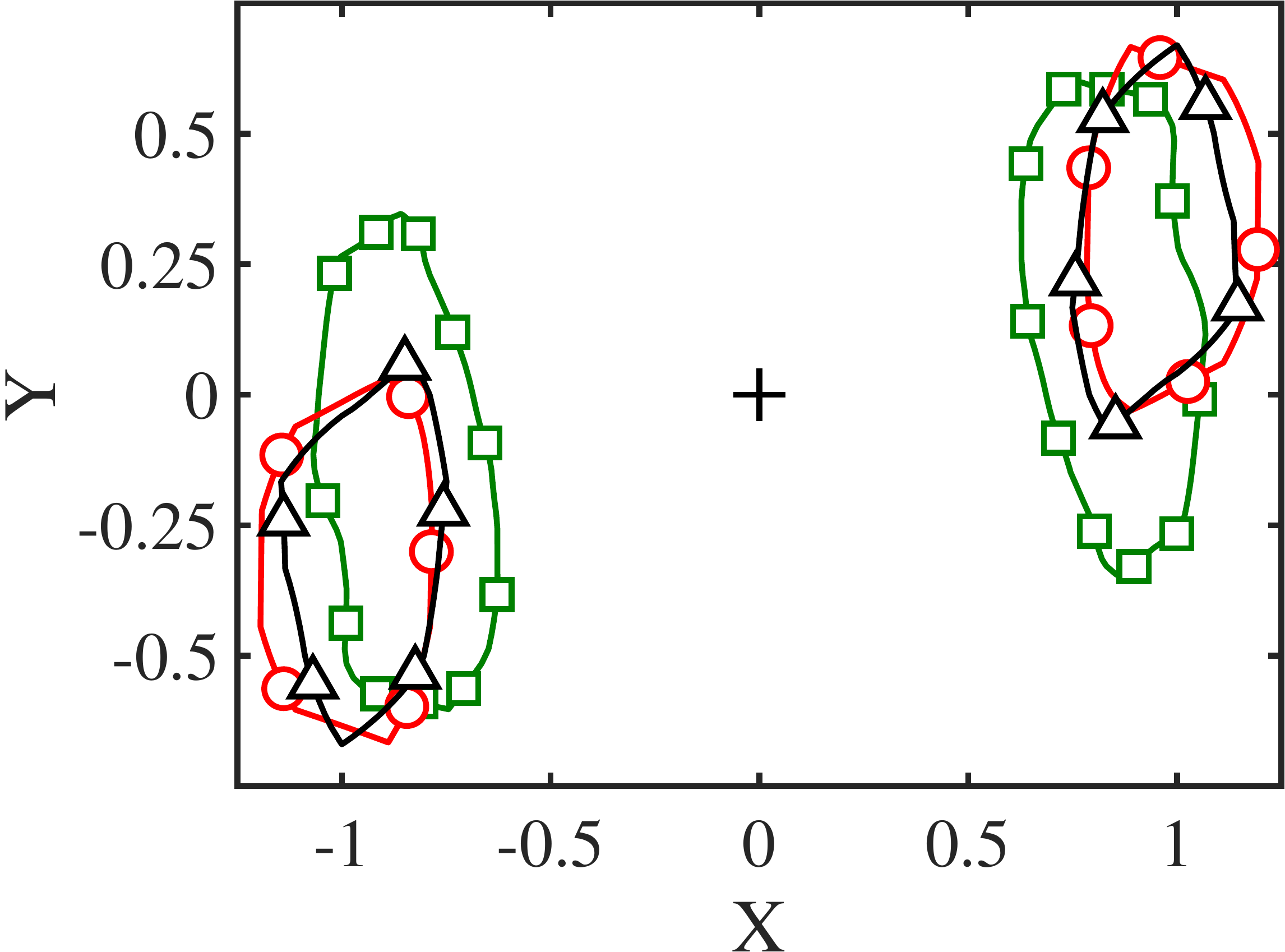}%
		\mylab{-4.5cm}{4cm}{(\bbb)}
}
	\caption{ Cross-sections of the three-dimensional joint
          p.d.f. of relative distances between the $\varPi$-structures
          for (a) $p^{\varPi^+ \varPi^-}$ in the ($\delta_z,\delta_y$)
          plane and (b) $p^{\varPi^+ \varPi^+}$ in the
          ($\delta_x,\delta_y$) plane. P.d.f.s are integrated over
          $\delta$=$\pm$0.2 normal to the plane of the plot. Contours
          contain the highest 15\% of the data. In all the panels, the
          symbols are $\square$, $r_f=23.3\eta$; $\triangle$,
          $r_f=46\eta$; $\bigcirc$, $r_f=60\eta$; and $+$,
          $\delta_x=\delta_y=\delta_z=0$. }
	\label{fig:pdf_rela_posi}
\end{figure}

Structures of $\varPi^+$ and $\varPi^-$ are preferentially organised
size-by-size in the spanwise direction, as shown by the cross-section
of $p^{\varPi^+ \varPi^-}$ in figure
\ref{fig:pdf_rela_posi}(a). \textcolor{black}{Except for
  $r_f=23.3\eta$, the} distribution of $p^{\varPi^+ \varPi^-}$ is
bi-modal along the vertical direction, with peaks lying almost
symmetrically at $\delta_y\approx \pm 0.25$.
%
Conversely, structures of the same type are aligned in the streamwise
direction with a separation of $|\delta_x| \approx 1$ and tilted by
roughly $15^\circ$. The p.d.f.s in figure \ref{fig:pdf_rela_posi}(b) are
for $p^{\varPi^+\varPi^+}$ at $\delta_z=0$, but similar results are found
for $p^{\varPi^-\varPi^-}$.

\begin{figure}
\vspace{0.5cm}
    \centerline{
    \includegraphics[width=0.98\textwidth]{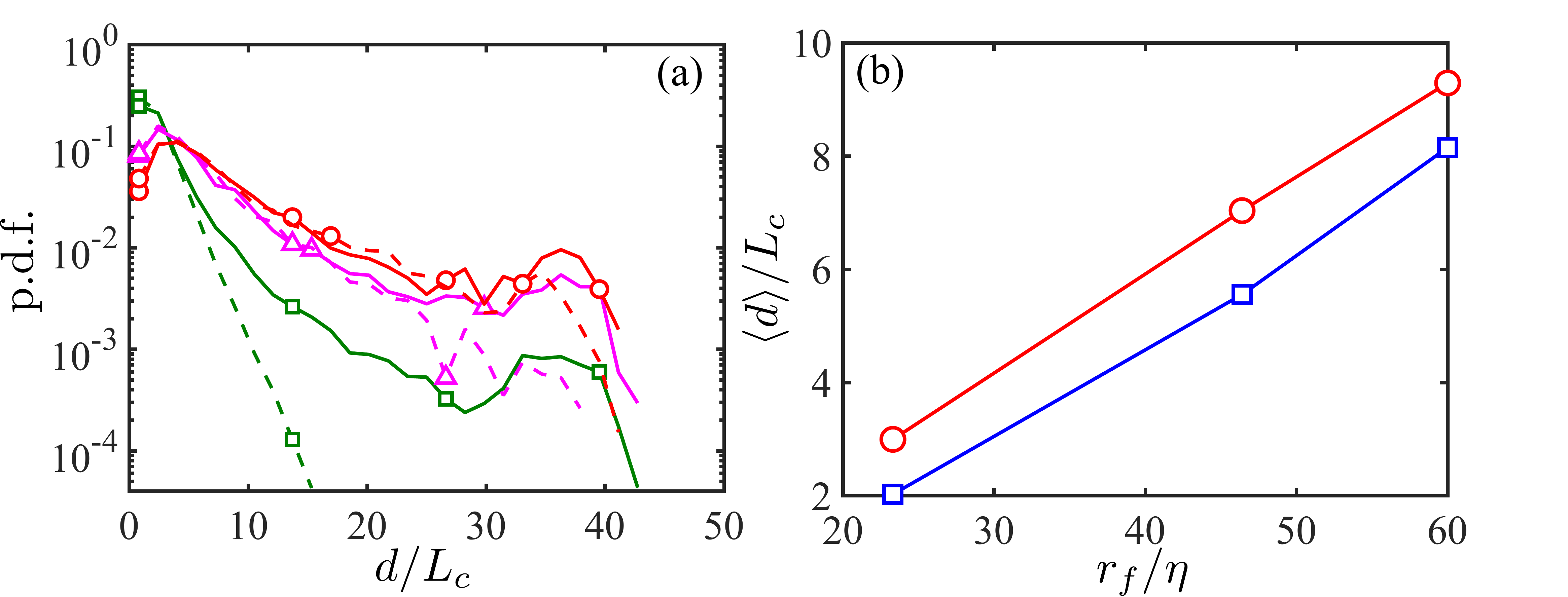}
}
	\caption{ \textcolor{black}{(a) P.d.f. of the diagonal length
            of individual three-dimensional $\varPi$ structures for
            $r_f=23.3\eta$ ($\square$), $r_f=46\eta$ ($\triangle$),
            and $r_f=60\eta$ ($\bigcirc$). Solid and dashed lines are
            for $\varPi^+$ and $\varPi^-$, respectively. (b) Mean
            value of the diagonal length of individual
            three-dimensional $\varPi$ structures as a function of the
            filter width. Colours and symbols are red circles for
            $\varPi^+$ and blue squares for $\varPi^-$.}}
	\label{fig:pdf_d_flux}
\end{figure}

\textcolor{black}{The p.d.f. and mean values of the diagonal length of
  individual $\varPi$-structures for different filter widths
  are plotted in figure \ref{fig:pdf_d_flux}(a) and
  figure \ref{fig:pdf_d_flux}(b), respectively. The results shows that
  the size of the structures are, on average, proportional to the
  filter width. Figure \ref{fig:pdf_d_flux}(a) shows some disparity
  between the p.d.f. of $d/L_c$ for $\varPi^+$ and $\varPi^-$, but the
  difference decreases with $r_f$, consistent with the skewness of the
  p.d.f. of $\varPi$ from figure \ref{fig:flux}(c).  For $\varPi^+$
  (similar trend for $\varPi^-$), the average values of $d/L_c$ are
  3.2, 7.3, and 9.7 for increasing $r_f$.  Therefore, the fair
  collapse of the contours in figure \ref{fig:pdf_rela_posi}(a) for
  $r_f>23.3\eta$ implies that the spatial organisation of $\varPi^+$
  and $\varPi^-$ is self-similar across the flow scales above the
  viscous range}.

\subsection{Coherent flow associated with the energy cascades}
\label{sec:coherent}

Once we have established the self-similar organisation of the cascades
in space, we characterise the three-dimensional flow conditioned to
the presence of $\varPi$-structures.  We follow the methodology of
\citet{don:lon:seki:jim:2017:jfm}; i.e., the flow is averaged in a
rectangular domain whose centre coincides with the centre of gravity
of the $n$-th structure, $\xvec_c^{n}$, and its edges are
$\boldsymbol{r}=(r_x,r_y,r_z)$ times the diagonal length $d^n$ of the
bounding box of the structure. The conditionally averaged quantity
$\widetilde{\phi}$ is then computed as
\beq
\{\widetilde{\phi}\}(\boldsymbol{r}) = \sum_{n=1}^{N} \frac{\widetilde{\phi}(\xvec_c^{n} + d^n \boldsymbol{r})}{N},
\la{eq:condave}
\eeq
where $n=1,..,N$ is the set of $\varPi$-structures selected for the
conditional average. \textcolor{black}{In the remainder of this work, the results are
for $r_f = 60\eta$, but similar conclusions are drawn for $r_f = 23.3\eta$
and $r_f = 46.6\eta$.} Additionally, we focus on the energy-transfer mechanism
for $\varPi$-structures with sizes larger than the Corrsin
length-scale, $d>L_c$, above which the mean shear dominates.

\begin{figure}
\vspace{0.5cm}
	\centerline{
\includegraphics[width=0.95\textwidth]{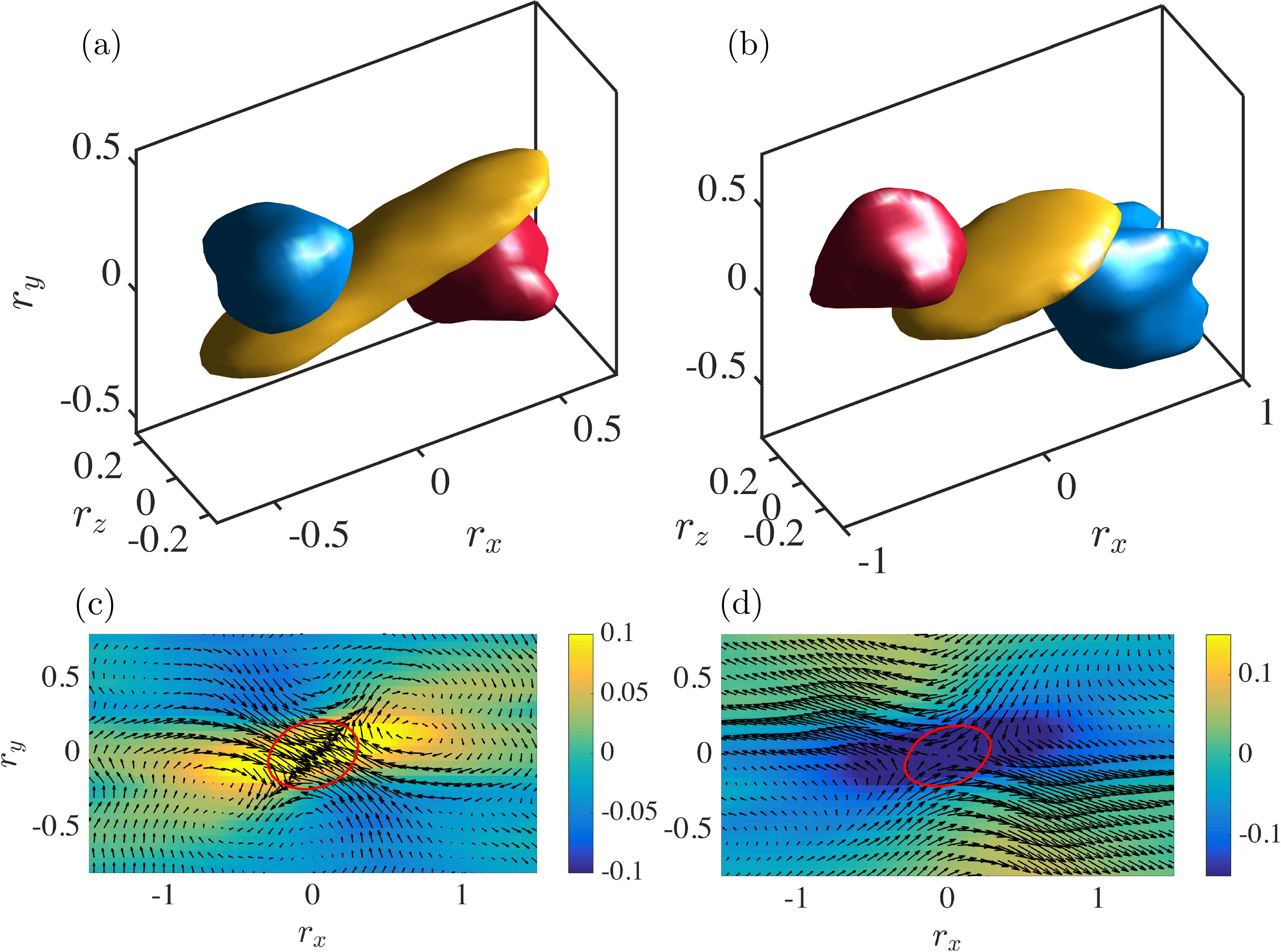}
	}
	\caption{ Averaged flow fields conditioned on
          $\varPi^+$-structures (a,c), and
          $\varPi^-$-structures (b,d). The
          isosurfaces in (a) and (\textcolor{black}{b}) are regions
          where $-\{\widetilde{u}\} \cdot\{\widetilde{v}\}$ and
          $\{\partial_y {\widetilde{u}}\}$ are larger than 0.4 and 0.7
          of their maximum values in the domain, respectively. The
          colours are blue for the sweep, red for the ejection, and
          yellow for the shear layer.  In panels
          (\textcolor{black}{c}) and (d), the arrows are
          $(\{\widetilde{u}\}, \{\widetilde{v}\})$, and the colours
          represent $\{\partial_y
          {\widetilde{u}}\}$. \textcolor{black}{Velocities are
            normalised by $SL_z$, $\partial \widetilde{u}/\partial y$ is
            normalised by the mean shear rate $S$, and distances are
            normalised by the diagonal length of individual structures
            as seen in (\ref{eq:condave})}. The solid
            red lines in panels (c) and (d) represent 0.6 of the
            maximum probability of finding a point belonging to a
            $\varPi^+$ or $\varPi^-$ structure, respectively.}
	\label{fig:cond_v}
\end{figure}

\subsubsection{\textcolor{black}{Averaged flow field conditioned on intense $\varPi$}}
\label{sec:cond_fu}

The three-dimensional velocities conditioned on
intense structures of $\varPi^+$ and $\varPi^-$ are
disclosed in figure \ref{fig:cond_v}(a,c) and (b,d),
  respectively. Panels (a) and (b) show the averaged tangential
Reynolds stress $\{\widetilde{u}\}\cdot \{\widetilde{v}\}$ and shear
layer $\{\partial_y {\widetilde{u}}\}$. To facilitate the
visualisation of the velocity field, figures \ref{fig:cond_v}(c) and
(d) contain the vector field $(\{\widetilde{u}\}, \{\widetilde{v}\})$
overlaid with $\{\partial_y {\widetilde{u}}\}$ in the plane $r_z=0$.
Regions of intense $\varPi$ are closely associated with a sweep
($\{\widetilde{u}\}>$0, $\{\widetilde{v}\}<$0) and an ejection
($\{\widetilde{u}\}<$0, $\{\widetilde{v}\}>$0) represented by the
regions coloured in blue and red, respectively, in figures
\ref{fig:cond_v}(a) and (b). For the forward
cascade, the ejection is located downstream and beneath $\varPi^+$,
while the sweep occurs upstream and above $\varPi^+$. Conversely, the
positions of the sweep and the ejection are interchanged for intense
$\varPi^-$.

Interestingly, figure \ref{fig:cond_v}(c) shows that
the forward energy transfer is confined within a large-scale shear
layer (the yellow region in figure \ref{fig:cond_v}a) originated by
the collision of the sweep and the ejection. Inversely,
as shown in figure \ref{fig:cond_v}(d) the backward
energy transfer occurs within the saddle region induced by the
separation of the sweep and the ejection. \textcolor{black}{The shear
  layers are inclined with respect to the streamwise direction by
  roughly $16^{\circ}$, with a characteristic length of about two
  times the diagonal length of the associated $\varPi$ structure. The
  inclination angle is almost identical to that of the streamwise
  fluxes train shown in figure \ref{fig:pdf_rela_posi}(b).  The
  intensity of the shear layer around $\varPi$ is 10\% of $S$ owing to
  the lack of small scales motions, which are filtered out. The
  intensities of the averaged vertical velocity $\{\widetilde{v}\}$
  conditioned on $\varPi^+$ and $\varPi^-$ are almost identical, but
  the averaged streamwise velocity $\{\widetilde{u}\}$ conditioned on
  $\varPi^-$ is roughly twice of that conditioned on $\varPi^+$.}
\begin{figure}
	\centerline{
\includegraphics[width=0.9\textwidth]{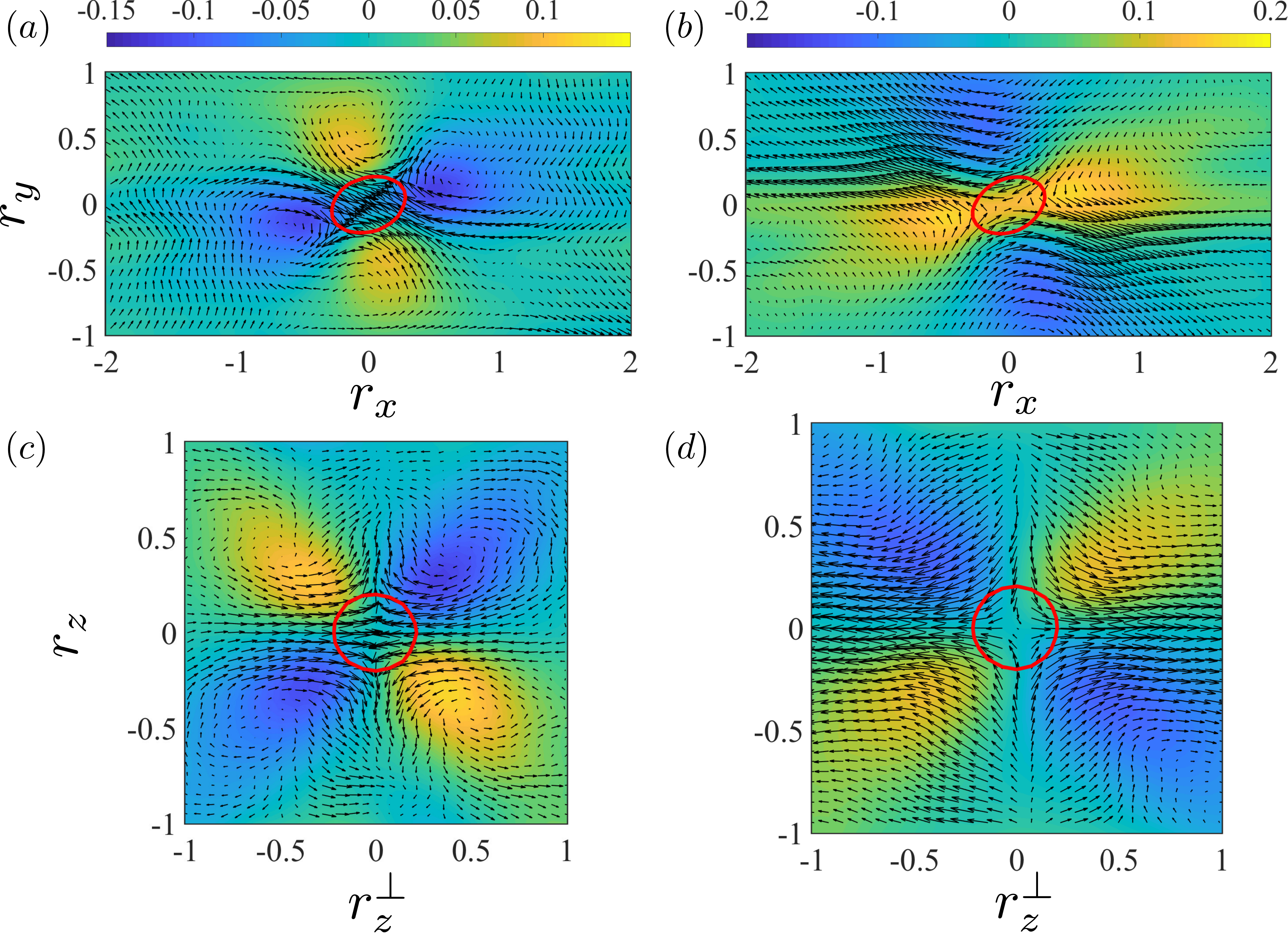}%
	}
	\caption{ \textcolor{black}{The averaged vorticity field
            conditioned on intense $\varPi^+$ (a,c) and $\varPi^-$
            (b,d).  Panels (a) and (b) are for spanwise vorticity $\{
            \widetilde{\omega_z} \}$ in the same plane as
            figure \ref{fig:cond_v}(a,b).  Panels
            (c-d) are for $(\{ \widetilde{\omega_x} \}$+$\{
            \widetilde{\omega_y} \})$cos$(45^\circ)$ in the plane
            containing $\deltavec=\textbf{0}$ with an inclination
            angle 135$^\circ$ with respect to the streamwise
            direction. $\delta_z^{\bot}=\delta_y/$cos$(45^\circ)$. Arrows
            represent $(\{\widetilde{u}\},\{\widetilde{v}\})$ in (a-d)
            and $(\{\widetilde{w}\}^{\bot},\{\widetilde{w}\})$ in
            (c,d), where $\{\widetilde{w}\}^{\bot}=(\{\widetilde{v}\}
            - \{\widetilde{u}\})$cos$(45^\circ)$. The range of the
            contour is $[-0.25\ 0.25]S$ in (a-b) and
            $[-0.125\ 0.125]S$ in (c-f) from blue to
            yellow. The solid red lines are the same
              as in figure \ref{fig:cond_v}}}
	\label{fig:cond_o}
\end{figure}

\textcolor{black}{The velocity patterns shown in figure
  \ref{fig:cond_v}, as well as the inclination angle of the shear
  layer, agree well with those in zero-pressure-gradient turbulent
  boundary layer \citep{natrajan:christensen:2006:pof}, in the channel
  with rough walls \citep{hong:katz:meneveau:schultz:2012:jfm} and in
  the atmosphere boundary layer with near-neutral atmospheric
  stability \citep{carper:portel:2004:jot}.  However, the organisation
  of the ejection and the sweep in SS-HST around $\varPi$ is symmetric
  due the nature of the flow, in contrast to the pattern observed in
  the near-wall region of wall-bounded turbulence
  \citep{pio:yu:adrian:1996, carper:portel:2004:jot}.}

\textcolor{black}{We analyse next the organisation of the vorticity
  $\{\widetilde{\omega_i}\}$ around intense energy transfer events. To
  gain a better insight into the flow organisation, it is convenient
  to represent the flow in the plane aligned with the mean inclination
  angle of the shear layer.  Figure \ref{fig:cond_o}(c-d) shows ($\{
  \widetilde{\omega_x} \}$+$\{\widetilde{\omega_y} \}$)cos$(45^\circ)$
  in the plane ($z^{\bot}-z$), which is inclined by
  135$^\circ$ with respect to the streamwise direction. The
  ($z^{\bot}-z$) plane contains $\deltavec=\textbf{0}$ and cuts
  through the centre of $\{\widetilde{\omega_x}\}$ and
  $\{\widetilde{\omega_y}\}$, with $z^{\bot}$ defined by
  $y/\textup{cos}(45^\circ)$. The velocity vectors in figure
  \ref{fig:cond_o}(c-d) are $(\{\widetilde{w}\}^{\bot},
  \{\widetilde{w}\})$ with $\{\widetilde{w}\}^{\bot}=
  (\{\widetilde{v}\}- \{\widetilde{u}\})$cos$(45^\circ)$. The
  inclination angles of the vorticity pairs with the same sign in
  $z^{\bot}-z$ plane are also $45^\circ$ or $135^\circ$.}

\textcolor{black}{Figure \ref{fig:cond_o}(a-b) show the conditional
  averaged spanwise vorticity, $\{\widetilde{\omega_z}\}$, in the same
  plane as in figure \ref{fig:cond_v}(c-d). The average spanwise
  vorticity shows a quadrupolar configuration: one pair of
  $\{\widetilde{\omega_z}\}$ is inclined by the same angle as the
  shear layer, whereas a weaker second pair appears with opposite sign
  normal to the shear layer. The result differs slightly from the
  observations in the channel flows with rough walls
  \citep{hong:katz:meneveau:schultz:2012:jfm}, where the
  $\{\widetilde{\omega_z}\}$ inclined with the shear layer does not
  form a pair, but a train in the downstream of the flux. The
  conditional averaged $\{\widetilde{\omega_x}\}$ and
  $\{\widetilde{\omega_y}\}$ also adopt a quadrupolar configuration,
  but different from the one obtained for
  $\{\widetilde{\omega_z}\}$. Both $\{\widetilde{\omega_x}\}$ and
  $\{\widetilde{\omega_y}\}$ are parallel to the $x-y$ plane and are
  inclined by $45^\circ$ with respect to the streamwise direction.}

\textcolor{black}{The quadrupolar $\{\widetilde{\omega_y}\}$ was also
  observed in the inner mixing zone of flow induced by the
  Richtmyer-Meshkov instability at the early stage of mixing process
  \citep{liu:xiao:2016:pre}, in the zero-pressure-gradient turbulent
  boundary layer \citep{carper:portel:2004:jot} and in the channel
  with rough walls \citep{hong:katz:meneveau:schultz:2012:jfm}.  In
  the latter two cases, the counter rotating
  $\{\widetilde{\omega_y}\}$ pairs in the downstream and upstream of
  the flux have different intensities, which reflects the different
  intensities of sweeps and ejections.}

\textcolor{black}{In summary, the flow patterns reported above suggest
  that the flow around $\varPi^+$ forms a saddle region contained in
  the streamwise and vertical direction such that $\{\partial_x
  \widetilde{u}\}<0$, $\{\partial_y \widetilde{v}\}<0$ and
  $\{\partial_z \widetilde{w}\}>0$.  The reversed pattern is observed
  for the flow around $\varPi^-$. Such abrupt changes would cause
  strong gradients and a significant kinetic energy transfer. The
  decomposition of $\varPi$ as the sum $\varPi = \sum_{ij}
  \varPi_{ij}$ with $\varPi_{ij} = -\tau_{ij} \partial \tilde u_i
  /\partial x_j$ reveals that, $\varPi_{11}$, $\varPi_{33}$ and
  $\varPi_{12}$ are the dominant terms, contributing 61.5\%, $-$30.0\%
  and 48.8\%, respectively, to the total $\varPi$.}

\subsubsection{Connection between conditional flow fields, vortex stretching,
    and strain self-amplification}
\label{subsec:VS}
  
\begin{figure}
\vspace{1cm}
	\centerline{
        \includegraphics[width=1\textwidth]{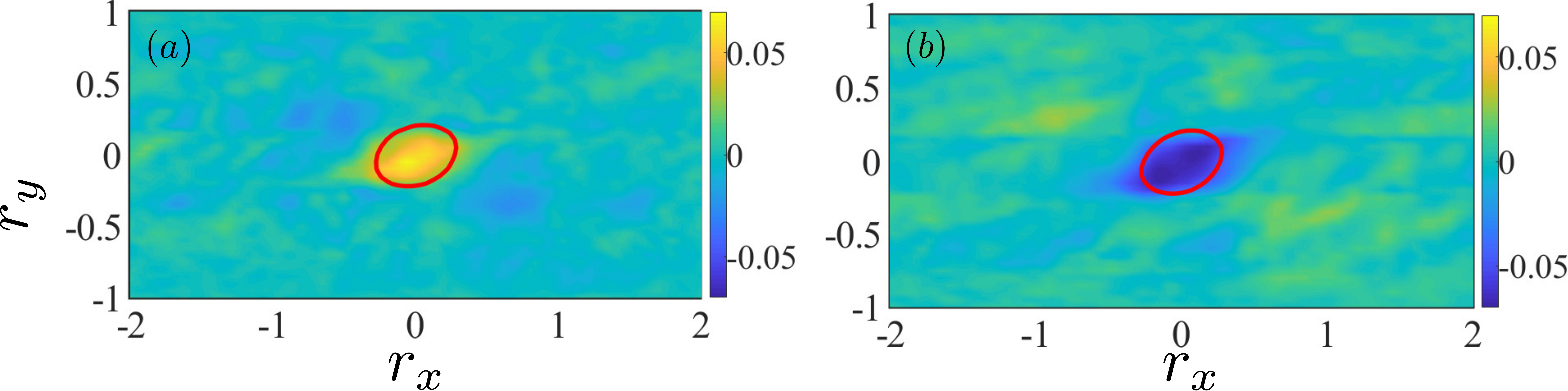}
	}
	\caption{The averaged $\beta$ (intermediate eigenvalue of
          $\boldsymbol{S}$) conditioned on intense $\varPi^+$ (a) and
          $\varPi^-$ (b). The values of $\beta$ are normalised by the
          mean shear $S$.}
	\label{fig:beta}
\end{figure}

\textcolor{black}{A vast body of literature places vortex stretching
  at the core of the energy transfer mechanism among scales
  \citep{Leung2012, Goto2017, LozanoHolznerJFM2016, Motoori2019},
  while recent works suggest that strain-rate
    self-amplification is the main contributor to the energy transfer
    \citep{car:bra:2019}. Thus, it is relevant to explore the
  connection between the flow structure identified in
  \S\ref{sec:cond_fu} and the vortex stretching and
    strain self-amplification mechanisms. Following
  \citet{Betchov1956}, the vortex stretching rate can be expressed as
    \begin{equation}
    \widetilde{\boldsymbol{\omega}}^{\top} \widetilde{\boldsymbol{S}}
    \widetilde{\boldsymbol{\omega}}=-4\textup{Tr}(\widetilde{\boldsymbol{S}}^3)=-12\alpha
    \beta \gamma,
    \end{equation}    
    where $\widetilde{\boldsymbol{S}}$ is the filtered rate of strain tensor, and
    $\alpha \ge \beta \ge \gamma$ are its real eigenvalues. The
    incompressibility condition requires that $\alpha + \beta + \gamma
    = 0$. The largest eigenvalue $\alpha$ is always positive
    (extensional), $\gamma$ is always negative (compressive), while
    $\beta$ can be either positive or negative depending on the
    magnitudes of $\alpha$ and $\gamma$.  Hence, the role played by
    vortex stretching is controlled by the sign of $\beta$.}
 \textcolor{black}{The values of $\beta$ conditioned on $\varPi^+$ and
   $\varPi^-$ are shown in figure \ref{fig:beta}. At the core of
   $\varPi^+$, $\beta>0$, which implies that, on average,
   $\widetilde{\boldsymbol{\omega}}^{\top} \widetilde{\boldsymbol{S}}
   \widetilde{\boldsymbol{\omega}}>0$.  The opposite is true at the
   core of $\varPi^-$, where $\widetilde{\boldsymbol{\omega}}^{\top}
   \widetilde{\boldsymbol{S}} \widetilde{\boldsymbol{\omega}}<0$ in
   the mean. The previous outcome suggests that vortex stretching is
   active during the forward cascade, while the vortex destruction
   dominates in regions of backward cascade.}
  Similarly, the strain self-amplification rate is
   \begin{equation}
   \widetilde{\boldsymbol{S}}^{\top} \widetilde{\boldsymbol{S}}
   \widetilde{\boldsymbol{S}}=-\frac{3}{4}\widetilde{\boldsymbol{\omega}}^{\top} \widetilde{\boldsymbol{S}}
   \widetilde{\boldsymbol{\omega}},
   \end{equation}   
   and the strain destruction ($\boldsymbol{S}^{\top} \boldsymbol{S}
   \boldsymbol{S}<0$) is bound to dominate in the regions of forward
   cascade, whereas the strain amplification ($\boldsymbol{S}^{\top}
   \boldsymbol{S} \boldsymbol{S}>0$) is active within backward cascade
   events.


\subsubsection{\textcolor{black}{Relation between intense $\varPi$ and hairpins}}

\begin{figure}
	\vspace{0.5cm}
	\centerline{
          \includegraphics[width=1\textwidth]{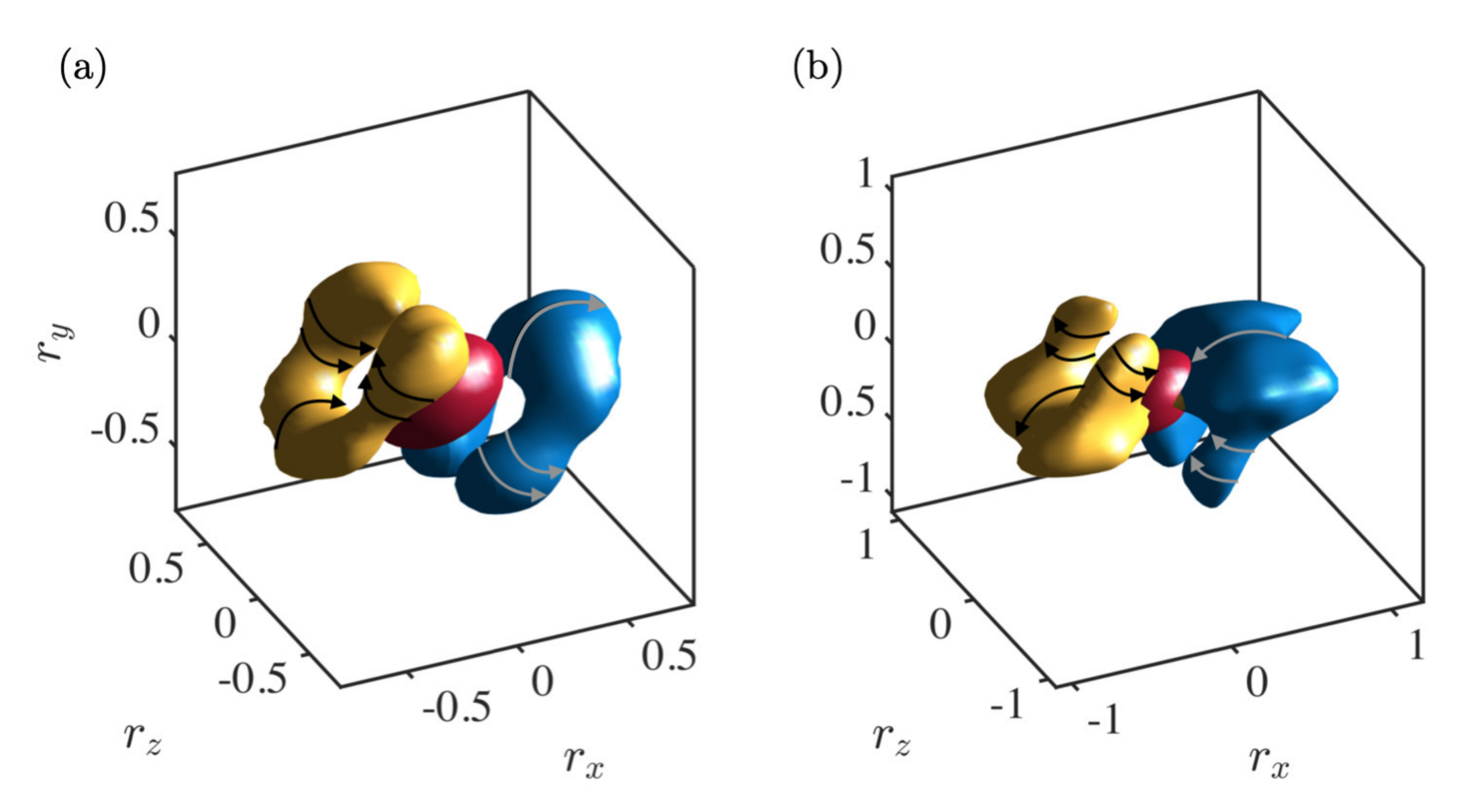}
}
	\caption{ Averaged conditional enstrophy $\{
          \widetilde{\omega} \}^2=\{ \widetilde{\omega_x} \}^2+\{
          \widetilde{\omega_y} \}^2+\{ \widetilde{\omega_z} \}^2$
          associated with forward (a) and backward (b) cascades.
          Upright and inverted hairpins, defined by $\{
          \widetilde{\omega} \}^2$=0.5$\{ \widetilde{\omega}
          \}^2_{\textup{max}}$, are coloured in blue and yellow,
          respectively. The isosurface coloured in red is 0.7 of the
          maximum probability of finding a point belonging to a
          $\varPi$-structure. The curved arrows indicate the direction
          of rotation of the flow \textcolor{black}{based on the
            signs of averaged vorticities.  In
            particular, the rotation of the legs is based on the
            signs of the averaged $\omega_x$ and
            $\omega_y$, while the rotation at the head is based on the
            averaged $\omega_z$.}} \label{fig:hairpin_flux}
\end{figure}

We inspect the enstrophy structure surrounding the forward and backward
energy cascades. To this end, we compute the averaged enstrophy
$\{\widetilde{\omega} \}^2=\{ \widetilde{\omega_x} \}^2+\{
\widetilde{\omega_y} \}^2+\{ \widetilde{\omega_z} \}^2$ conditioned on
structures of forward and backward kinetic energy transfer, where
$\omega_i$ is the vorticity. As seen in figure \ref{fig:hairpin_flux},
both $\varPi^+$ and $\varPi^-$ are located at the leading edge of an
upstream inverted hairpin and at the trailing edge of a downstream
upright hairpin. Given the average nature of the conditional flow, the
emerging upright and inverted hairpins should be appraised as
statistical manifestations rather than instantaneous features of the
flow.
\textcolor{black}{It was assessed that repeating the analysis using
  only half of the flow fields does not altered the conclusions
  presented above. More precisely, if we denote by $\{\tilde
  \omega\}_1$ and $\{\tilde \omega\}_2$ the average vorticity field
  obtained using all and half of the flow fields, respectively, the
  relative difference between both fields was found to be at most
  1\%. Similar values are obtained for other quantities.}

Our results show that both upright and inverted hairpins are involved
in the kinetic energy transfer. The flow representation promoted above
differs from previous models in which upright hairpins dominate
\citep{carper:portel:2004:jot,natrajan:christensen:2006:pof}.
\textcolor{black}{The sweep and ejection around the kinetic energy flux
  are attributed respectively to the head and legs of upright
  hairpins.}  Nonetheless, those studies were hampered by
two-dimensional observations, whereas we have highlighted that a fully
three-dimensional analysis is necessary to elucidate the actual
enstrophy structure of $\varPi$. \textcolor{black}{As supplementary
  material, we provide videos of the flow patterns in figure
  \ref{fig:hairpin_flux} and figures \ref{fig:cond_v}(a,b) to assist
  the reader in understanding the spatial structure of the energy
  cascade [movie\_S1,S2,S3, and S4].}
Individual inverted hairpins have been observed numerically and
experimentally in homogeneous shear turbulence
\citep{rog:moi:87,van:tav:11} and in channels flows
\citep{kim:moi:86}, and other investigations have also linked
$\varPi^+$ to U-shaped regions in the flow \citep{gerz:how:mah:1994,
  fin:shaw::pat:09, hong:katz:meneveau:schultz:2012:jfm} akin to the
inverted hairpin reported
here. \textcolor{black}{\cite{hong:katz:meneveau:schultz:2012:jfm}
  concluded that both the sweep and the ejection around $\varPi^+$ are
  induced by the legs hairpins aligned in the streamwise direction.
  Our results suggest, however, that both upright and inverted
  hairpins are involved in the generation of sweeps and ejections
  around $\varPi$.}

\section{Conclusions}\label{sec:conclude}

We have studied the three-dimensional flow structure and organisation
of the kinetic energy transfer in shear turbulence for flow scales
above the Corrsin length. Our analysis is focused on spatially
intermittent regions where the transfer of energy among flow scales is
intense. The structure of the velocity and enstrophy fields around
these regions has been investigated separately for forward cascade
events and backward cascade events.

The inspection of the relative distances between forward and backward
cascades has shown that positive and negative energy transfers are
paired in the spanwise direction and that such pairs form a train
aligned in the mean-flow direction. Our results also indicate that the
latter arrangement is self-similar across flow scales.  We have
further uncovered that the energy transfer is accompanied by nearby
upright and inverted hairpins, and that the forward and backward
cascades occur, respectively, within the shear layer and the saddle
region lying in between the hairpins. The asymmetry between the
forward and backward cascades emerges from the opposite flow
circulation of the associated upright/inverted hairpins, which prompts
reversed patterns in the flow. To date, the present findings represent
the most detailed structural description of the energy cascade in
shear turbulence.  In virtue of the previously reported similarities
between wall turbulence and SS-HST, we expect our results to be
representative of wall-bounded flows. Nonetheless, additional efforts
should be devoted to confirm this scenario in other flow
configurations.

We have applied our method to the simplest shear turbulence, but
nothing prevents its application to the inertial range of isotropic
turbulence and to more complex configurations with rotation, heat
transfer, electric fields, quantum effects, etc. Finally, the
characterisation of the cascade presented in this study is static and
do not contain dynamic information on how the kinetic energy is
transferred in time between hairpins at different scales. Therefore,
our work is just the starting point for future investigations and
opens new venues for the analysis of the time-resolved structure of
the energy cascade \citep{Cardesa2017}, including cause-effect
interactions among energy-containing eddies \citep{Lozano2020}.

\begin{acknowledgements}
This work was supported by the National Key Research and Development
Program of China (2016YFA0401200), the European Research Council
(ERC-2010.AdG-20100224, ERC-2014.AdG-669505), the National Science
Foundation of China (11702307, 11732010), Fundamental Research Funds
for the Central Universities (20720180120), and MEL Internal Research
Fund (MELRI1802). The authors would like to acknowledge fruitful
discussions with Prof. Jim\'enez and Dr. C. Pan, BUAA.
\end{acknowledgements}

\bibliographystyle{jfm}
\bibliography{flux_rapids}

\end{document}

%% file: macros.tex
\newcommand{\itbold}[1]{\textbf{\textit{#1}}}

\newcommand{\pdif}[2]{\dfrac{\partial #1}{\partial #2}}
\newcommand{\pddif}[2]{\dfrac{\partial^2 #1}{\partial {#2}^2}}

\newcommand{\uvec}{\itbold{u}}

\newcommand{\xvec}{\itbold{x}}
\newcommand{\Hvec}{\itbold{H}}
  
\newcommand{\omgvec}{\boldsymbol{\omega}}

\def\bra{\langle}
\def\ket{\rangle}

\def\beq{\begin{equation}}
\def\eeq{\end{equation}}
\def\la{\label}

\def\ii{{\rm i}}

\def\bra{\langle}
\def\ket{\rangle}

\def\beq{\begin{equation}}
\def\eeq{\end{equation}}
\def\la{\label}
\def\ii{{\rm i}}

\def\utau{u_\tau}

\def\xvec{\boldsymbol{x}}

\def\deltavec{\boldsymbol{\delta}}
\definecolor{olivegreen}{rgb}{0,0.6,0}

\newcommand{\mylab}[3]{\raisebox{#2}[0mm][0mm]{%
\makebox[0mm][l]{\hspace*{#1}\textbf{#3}}}}
\def\spacce#1{\hskip #1pt}
\def\drawline#1#2{\raise 2.5pt\vbox{\hrule width #1pt height #2pt}}
\def\solid{\drawline{24}{.5}\nobreak}

\def\bdash{\hbox{\drawline{5.8}{.5}\spacce{2}}}

\def\dashed{\bdash\bdash\bdash\nobreak}

\def\chndot{\hbox%
{\drawline{4.6}{.5}\spacce{2}\drawline{1}{.5}\spacce{2}\drawline{4.6}{.5}\spacce{2}\drawline{1}{.5}\spacce{2}\drawline{4.6}{.5}}\nobreak }

\def\trian{\raise 1.25pt\hbox{$\scriptstyle\triangle$}\nobreak}

\def\dtrian{\raise 1.25pt\hbox%
{$\scriptscriptstyle\bigtriangledown$}\nobreak}

\def\squar{\raise 1.25pt\hbox{$\scriptstyle\Box$}\nobreak}

\def\diamon{\raise 1.25pt\hbox{$\scriptstyle\diamond$}\nobreak}

\def\bra{\langle}
\def\ket{\rangle}

\def\beq{\begin{equation}}
\def\eeq{\end{equation}}
\def\la{\label}
\def\utau{u_\tau}

\def\aaa{{\it a}}
\def\bbb{{\it b}}

\def\citalajim03{Del \'Alamo \& Jim\'enez (2003)}



%% file: flux_rapids.bbl
\begin{thebibliography}{61}
\expandafter\ifx\csname natexlab\endcsname\relax\def\natexlab#1{#1}\fi
\def\au#1{#1} \def\ed#1{#1} \def\yr#1{#1}\def\at#1{#1}\def\jt#1{\textit{#1}}
  \def\bt#1{#1}\def\bvol#1{\textbf{#1}} \def\vol#1{#1} \def\pg#1{#1}
  \def\publ#1{#1}\def\arxiv#1{#1}\def\org#1{#1}\def\st#1{\textit{#1}}

\bibitem[Alexakis \& Biferale(2018)]{Alexakis2018}
{\sc \au{Alexakis, A.} \& \au{Biferale, L.}} \yr{2018}  \at{Cascades and
  transitions in turbulent flows}.  \jt{Phys. Rep.}  \bvol{767-769},
  \pg{1--101}.

\bibitem[Aoyama {\em et~al.\/}(2005)Aoyama, Ishihara, Kaneda, Yokokawa, Itakura
  \& Uno]{aoyama:ishihara:2005}
{\sc \au{Aoyama, T.}, \au{Ishihara, T.}, \au{Kaneda, Y.}, \au{Yokokawa, M.},
  \au{Itakura, K.} \& \au{Uno, A.}} \yr{2005}  \at{Statistics of energy
  transfer in high-resolution direct numerical simulation of turbulence in a
  periodic box}.  \jt{J. Phys. Soc. Jpn.}  \bvol{74},  \pg{3202--3212}.

\bibitem[Balbus \& Hawley(1998)]{BalbusHawley1998}
{\sc \au{Balbus, S.~A.} \& \au{Hawley, J.~F.}} \yr{1998}  \at{Instability,
  turbulence, and enhanced transport in accretion disks}.  \jt{Rev. Mod. Phys.}
   \bvol{70},  \pg{1--53}.

\bibitem[Ballouz \& Ouellette(2018)]{Ballouz2018}
{\sc \au{Ballouz, Joseph~G.} \& \au{Ouellette, Nicholas~T.}} \yr{2018}
  \at{Tensor geometry in the turbulent cascade}.  \jt{J. Fluid Mech.}
  \bvol{835},  \pg{1048--1064}.

\bibitem[Baron(1982)]{Baron1982}
{\sc \au{Baron, F.}} \yr{1982}  \at{Macro-simulation tridimensionelle
  d{'\'e}coulements turbulents cisaill{\'e}s}. PhD thesis, U. Pierre et Marie
  Curie.

\bibitem[Betchov(1956)]{Betchov1956}
{\sc \au{Betchov, R.}} \yr{1956}  \at{An inequality concerning the production
  of vorticity in isotropic turbulence}.  \jt{J. Fluid Mech.}  \bvol{1},
  \pg{497--504}.

\bibitem[Bodenschatz(2015)]{Bodenschatz2015}
{\sc \au{Bodenschatz, E.}} \yr{2015}  \at{Clouds resolved}.  \jt{Science}
  \bvol{350}~(6256),  \pg{40--41}.

\bibitem[Bose \& Park(2018)]{Bose2018}
{\sc \au{Bose, S.~T.} \& \au{Park, G.~I.}} \yr{2018}  \at{Wall-modeled
  large-eddy simulation for complex turbulent flows}.  \jt{Annu. Rev. Fluid
  Mech.}  \bvol{50}~(1),  \pg{535--561}.

\bibitem[Carbone \& Bragg(2019)]{car:bra:2019}
{\sc \au{Carbone, M.} \& \au{Bragg, A.~D.}} \yr{2019}  \at{Is vortex stretching
  the main cause of the turbulent energy cascade?}  \jt{J. Fluid Mech.}
  \bvol{883},  \pg{R2, 1--13}.

\bibitem[Cardesa {\em et~al.\/}(2017)Cardesa, Vela-Mart{\'\i}n \&
  Jim{\'e}nez]{Cardesa2017}
{\sc \au{Cardesa, J.~I.}, \au{Vela-Mart{\'\i}n, A.} \& \au{Jim{\'e}nez, J.}}
  \yr{2017}  \at{The turbulent cascade in five dimensions}.  \jt{Science}
  \bvol{357}~(6353),  \pg{782--784}.

\bibitem[Carper \& Port{\'{e}}-Agel(2004)]{carper:portel:2004:jot}
{\sc \au{Carper, M.~A.} \& \au{Port{\'{e}}-Agel, F.}} \yr{2004}  \at{{The role
  of coherent structures in subfilter-scale dissipation of turbulence measured
  in the atmospheric surface layer}}.  \jt{J. Turbulence}  \bvol{5}~(04),
  \pg{1--24}.

\bibitem[Cerutti \& Meneveau(1998)]{Cerutti1998}
{\sc \au{Cerutti, Stefano} \& \au{Meneveau, Charles}} \yr{1998}
  \at{Intermittency and relative scaling of subgrid-scale energy dissipation in
  isotropic turbulence}.  \jt{Phy. Fluids}  \bvol{10}~(4),  \pg{928--937}.

\bibitem[Champagne {\em et~al.\/}(1970)Champagne, Harris \&
  Corrsin]{Champagne1970}
{\sc \au{Champagne, F.~H.}, \au{Harris, V.~G.} \& \au{Corrsin, S.}} \yr{1970}
  \at{Experiments on nearly homogeneous turbulent shear flow}.  \jt{J. Fluid
  Mech.}  \bvol{41}~(1),  \pg{81--139}.

\bibitem[Del~{\'A}lamo {\em et~al.\/}(2004)Del~{\'A}lamo, Jim{\'e}nez,
  Zandonade \& Moser]{del:jim:zan:mos:06}
{\sc \au{Del~{\'A}lamo, J.~C.}, \au{Jim{\'e}nez, J.}, \au{Zandonade, P.} \&
  \au{Moser, R.~D.}} \yr{2004}  \at{{Self-similar vortex clusters in the
  turbulent logarithmic region}}.  \jt{J. Fluid Mech.}  \bvol{561},
  \pg{329--358}.

\bibitem[Domaradzki {\em et~al.\/}(1993)Domaradzki, Liu \&
  Brachet]{Domaradzki1993}
{\sc \au{Domaradzki, J.~Andrzej}, \au{Liu, Wei} \& \au{Brachet, Marc~E.}}
  \yr{1993}  \at{An analysis of subgrid-scale interactions in numerically
  simulated isotropic turbulence}.  \jt{Phys. Fluids}  \bvol{5}~(7),
  \pg{1747--1759}.

\bibitem[Dong {\em et~al.\/}(2017)Dong, Lozano-Dur{\'{a}}n, Sekimoto \&
  Jim{\'{e}}nez]{don:lon:seki:jim:2017:jfm}
{\sc \au{Dong, S.}, \au{Lozano-Dur{\'{a}}n, A.}, \au{Sekimoto, A.} \&
  \au{Jim{\'{e}}nez, J.}} \yr{2017}  \at{{Coherent structures in statistically
  stationary homogeneous shear turbulence}}.  \jt{J. Fluid Mech.}  \bvol{816},
  \pg{167--208}.

\bibitem[Dubrulle(2019)]{dubrulle2019}
{\sc \au{Dubrulle, B\'ereng\`ere}} \yr{2019}  \at{Beyond kolmogorov cascades}.
  \jt{J. Fluid Mech.}  \bvol{867},  \pg{P1}.

\bibitem[Falkovich(2009)]{Falkovich2009}
{\sc \au{Falkovich, Gregory}} \yr{2009}  \at{Symmetries of the turbulent
  state}.  \jt{J. Phys. A}  \bvol{42}~(12),  \pg{123001}.

\bibitem[Finnigan {\em et~al.\/}(2009)Finnigan, Shaw \&
  Patton]{fin:shaw::pat:09}
{\sc \au{Finnigan, J.~J.}, \au{Shaw, R.~H.} \& \au{Patton, E.~G.}} \yr{2009}
  \at{{Turbulence structure above a vegetation canopy}}.  \jt{J. Fluid Mech.}
  \bvol{637},  \pg{387--424}.

\bibitem[Gerz {\em et~al.\/}(1994)Gerz, Howell \& Mahrt]{gerz:how:mah:1994}
{\sc \au{Gerz, T.}, \au{Howell, J.} \& \au{Mahrt, L.}} \yr{1994}  \at{Vortex
  structures and microfronts}.  \jt{Phys. Fluids}  \bvol{6}~(3),
  \pg{1242--1251}.

\bibitem[Gerz {\em et~al.\/}(1989)Gerz, Schumann \&
  Elghobashi]{GerzSchumannElghobashi1988}
{\sc \au{Gerz, T.}, \au{Schumann, U.} \& \au{Elghobashi, S.~E.}} \yr{1989}
  \at{Direct numerical simulation of stratified homogeneous turbulent shear
  flows}.  \jt{J. Fluid Mech.}  \bvol{200},  \pg{563--594}.

\bibitem[Goto {\em et~al.\/}(2017)Goto, Saito \& Kawahara]{Goto2017}
{\sc \au{Goto, S.}, \au{Saito, Y.} \& \au{Kawahara, G.}} \yr{2017}
  \at{Hierarchy of antiparallel vortex tubes in spatially periodic turbulence
  at high reynolds numbers}.  \jt{Phys. Rev. Fluids}  \bvol{2},  \pg{064603}.

\bibitem[H{\"{a}}rtel {\em et~al.\/}(1994)H{\"{a}}rtel, Kleiser, Unger \&
  Friedrich]{har:kle:ung:fri:1994:pof}
{\sc \au{H{\"{a}}rtel, C.}, \au{Kleiser, L.}, \au{Unger, F.} \& \au{Friedrich,
  R.}} \yr{1994}  \at{{Subgrid-scale energy transfer in the near-wall region of
  turbulent flows}}.  \jt{Phys. Fluids}  \bvol{6}~(9),  \pg{3130--3143}.

\bibitem[Hof {\em et~al.\/}(2010)Hof, De~Lozar, Avila, Tu \&
  Schneider]{Hof2010}
{\sc \au{Hof, B.}, \au{De~Lozar, A.}, \au{Avila, M.}, \au{Tu, X.} \&
  \au{Schneider, T.~M.}} \yr{2010}  \at{Eliminating turbulence in spatially
  intermittent flows}.  \jt{Science}  \bvol{327}~(5972),  \pg{1491--1494}.

\bibitem[Hong {\em et~al.\/}(2012)Hong, Katz, Meneveau \&
  Schultz]{hong:katz:meneveau:schultz:2012:jfm}
{\sc \au{Hong, J.}, \au{Katz, J.}, \au{Meneveau, C.} \& \au{Schultz, M.~P.}}
  \yr{2012}  \at{{Coherent structures and associated subgrid-scale energy
  transfer in a rough-wall turbulent channel flow}}.  \jt{J. Fluid Mech.}
  \bvol{712},  \pg{92--128}.

\bibitem[Ishihara {\em et~al.\/}(2009)Ishihara, Gotoh \& Kaneda]{Ishihara2009}
{\sc \au{Ishihara, Takashi}, \au{Gotoh, Toshiyuki} \& \au{Kaneda, Yukio}}
  \yr{2009}  \at{Study of high-reynolds number isotropic turbulence by direct
  numerical simulation}.  \jt{Annu. Rev. Fluid Mech.}  \bvol{41}~(1),
  \pg{165--180}.

\bibitem[Kawata \& Alfredsson(2018)]{Kawata2018}
{\sc \au{Kawata, T.} \& \au{Alfredsson, P.~H.}} \yr{2018}  \at{Inverse
  interscale transport of the reynolds shear stress in plane couette
  turbulence}.  \jt{Phys. Rev. Lett.}  \bvol{120},  \pg{244501}.

\bibitem[Kim \& Moin(1986)]{kim:moi:86}
{\sc \au{Kim, J.} \& \au{Moin, P.}} \yr{1986}  \at{{The structure of the
  vorticity field in turbulent channel flow. Part 2. Study of ensemble-averaged
  fields}}.  \jt{J. Fluid Mech.}  \bvol{162},  \pg{339--363}.

\bibitem[Kim \& Moin(1987)]{rog:moi:87}
{\sc \au{Kim, J.} \& \au{Moin, P.}} \yr{1987}  \at{{The structure of the
  vorticity field in homogeneous turbulent flows}}.  \jt{J. Fluid Mech.}
  \bvol{176},  \pg{33--66}.

\bibitem[Kim {\em et~al.\/}(1987)Kim, Moin \& Moser]{KimMoinMoser1987}
{\sc \au{Kim, J.}, \au{Moin, P.} \& \au{Moser, R.~D.}} \yr{1987}  \at{Turbulent
  statistics in fully developed channel flow at low {R}eynolds number.}  \jt{J.
  Fluid Mech.}  \bvol{177},  \pg{133--166}.

\bibitem[Kolmogorov(1941)]{Kolmogorov1941}
{\sc \au{Kolmogorov, A.~N.}} \yr{1941} {The local structure of turbulence in
  incompressible viscous fluid for very large {R}eynolds' numbers}.  \bt{In
  {\em Dokl. Akad. Nauk SSSR\/}}, ,  \vol{vol.~30},  \pg{pp. 301--305}.

\bibitem[Kolmogorov(1962)]{Kolmogorov1962}
{\sc \au{Kolmogorov, A.~N.}} \yr{1962}  \at{A refinement of previous hypotheses
  concerning the local structure of turbulence in a viscous incompressible
  fluid at high reynolds number}.  \jt{J. Fluid Mech.}  \bvol{13}~(1),
  \pg{82--85}.

\bibitem[K{\"u}hnen {\em et~al.\/}(2018)K{\"u}hnen, Song, Scarselli, Budanur,
  Riedl, Willis, Avila \& Hof]{Kuhnen2018}
{\sc \au{K{\"u}hnen, J.}, \au{Song, B.}, \au{Scarselli, D.}, \au{Budanur,
  N.~B.}, \au{Riedl, M.}, \au{Willis, A.~P.}, \au{Avila, M.} \& \au{Hof, B.}}
  \yr{2018}  \at{Destabilizing turbulence in pipe flow}.  \jt{Nat. Phys.}
  \bvol{14}~(4),  \pg{386--390}.

\bibitem[Leung {\em et~al.\/}(2012)Leung, Swaminathan \& Davidson]{Leung2012}
{\sc \au{Leung, T.}, \au{Swaminathan, N.} \& \au{Davidson, P.~A.}} \yr{2012}
  \at{Geometry and interaction of structures in homogeneous isotropic
  turbulence}.  \jt{J. Fluid Mech.}  \bvol{710},  \pg{453--481}.

\bibitem[Lin(1999)]{lin:1999:pof}
{\sc \au{Lin, C.}} \yr{1999}  \at{{Near-grid-scale energy transfer and coherent
  structures in the convective planetary boundary layer}}.  \jt{Phys. Fluids}
  \bvol{11}~(11),  \pg{3482--3494}.

\bibitem[Liu \& Xiao(2016)]{liu:xiao:2016:pre}
{\sc \au{Liu, H.} \& \au{Xiao, Z.}} \yr{2016}  \at{{Scale-to-scale energy
  transfer in mixing flow induced by the Richtmyer-Meshkov instability}}.
  \jt{Phys. Rev. E}  \bvol{93}~(5),  \pg{1--15}.

\bibitem[Lozano-Dur\'an {\em et~al.\/}(2020)Lozano-Dur\'an, Bae \&
  Encinar]{Lozano2020}
{\sc \au{Lozano-Dur\'an, Adri\'an}, \au{Bae, H.~Jane} \& \au{Encinar,
  Miguel~P.}} \yr{2020}  \at{Causality of energy-containing eddies in wall
  turbulence}.  \jt{J. Fluid Mech.}  \bvol{882},  \pg{A2}.

\bibitem[Lozano-Dur{\'{a}}n {\em et~al.\/}(2012)Lozano-Dur{\'{a}}n, Flores \&
  Jim{\'{e}}nez]{loz:flo:jim:12}
{\sc \au{Lozano-Dur{\'{a}}n, A.}, \au{Flores, O.} \& \au{Jim{\'{e}}nez, J.}}
  \yr{2012}  \at{{The three-dimensional structure of momentum transfer in
  turbulent channels}}.  \jt{J. Fluid Mech.}  \bvol{694},  \pg{100--130}.

\bibitem[Lozano-Dur{\'a}n {\em et~al.\/}(2016)Lozano-Dur{\'a}n, Holzner \&
  Jim{\'e}nez]{LozanoHolznerJFM2016}
{\sc \au{Lozano-Dur{\'a}n, A.}, \au{Holzner, M.} \& \au{Jim{\'e}nez, J.}}
  \yr{2016}  \at{Multiscale analysis of the topological invariants in the
  logarithmic region of turbulent channels at a friction {R}eynolds number of
  932}.  \jt{J. Fluid Mech.}  \bvol{803},  \pg{356--394}.

\bibitem[Lozano-Dur{\'a}n \& Jim{\'e}nez(2014)]{Lozano2014}
{\sc \au{Lozano-Dur{\'a}n, A.} \& \au{Jim{\'e}nez, J.}} \yr{2014}
  \at{Time-resolved evolution of coherent structures in turbulent channels:
  characterization of eddies and cascades}.  \jt{J. Fluid. Mech.}  \bvol{759},
  \pg{432--471}.

\bibitem[Marusic {\em et~al.\/}(2010)Marusic, Mathis \& Hutchins]{Marusic2010}
{\sc \au{Marusic, I.}, \au{Mathis, R.} \& \au{Hutchins, N.}} \yr{2010}
  \at{Predictive model for wall-bounded turbulent flow}.  \jt{Science}
  \bvol{329}~(5988),  \pg{193--196}.

\bibitem[Moisy \& Jim{\'e}nez(2004)]{moi:jim:04}
{\sc \au{Moisy, F.} \& \au{Jim{\'e}nez, J.}} \yr{2004}  \at{{Geometry and
  clustering of intense structures in isotropic turbulence}}.  \jt{J. Fluid
  Mech.}  \bvol{513},  \pg{111--133}.

\bibitem[Motoori \& Goto(2019)]{Motoori2019}
{\sc \au{Motoori, Yutaro} \& \au{Goto, Susumu}} \yr{2019}  \at{Generation
  mechanism of a hierarchy of vortices in a turbulent boundary layer}.  \jt{J.
  Fluid Mech.}  \bvol{865},  \pg{1085--1109}.

\bibitem[Natrajan \& Christensen(2006)]{natrajan:christensen:2006:pof}
{\sc \au{Natrajan, V.~K.} \& \au{Christensen, K.~T.}} \yr{2006}  \at{{The role
  of coherent structures in subgrid-scale energy transfer within the log layer
  of wall turbulence}}.  \jt{Phys. Fluids}  \bvol{18}~(6).

\bibitem[Obukhov(1941)]{Obukhov1941}
{\sc \au{Obukhov, AM}} \yr{1941}  \at{On the distribution of energy in the
  spectrum of turbulent flow}.  \jt{Bull. Acad. Sci. USSR, Geog. Geophys.}
  \bvol{5},  \pg{453--466}.

\bibitem[Osawa \& Jim{\'{e}}nez(2018)]{osawa:jimenez:2018}
{\sc \au{Osawa, K.} \& \au{Jim{\'{e}}nez, J.}} \yr{2018}  \at{{Intense
  structures of different momentum fluxes in turbulent channels}}.  \jt{Phys.
  Rev. Fluids}  \bvol{3}~(8),  \pg{1--14}.

\bibitem[Piomelli {\em et~al.\/}(1991)Piomelli, Cabot, Moin \&
  Lee]{pio:cab:moin:lee:1991}
{\sc \au{Piomelli, U.}, \au{Cabot, W.~H.}, \au{Moin, P.} \& \au{Lee, S.}}
  \yr{1991}  \at{{Subgrid-scale backscatter in turbulent and transitional
  flows}}.  \jt{Phys. Fluids}  \bvol{3}~(7),  \pg{1766--1771}.

\bibitem[Piomelli {\em et~al.\/}(1996)Piomelli, Yu \&
  Adrian]{pio:yu:adrian:1996}
{\sc \au{Piomelli, U.}, \au{Yu, Y.} \& \au{Adrian, R.~J.}} \yr{1996}
  \at{{Subgrid-scale energy transfer and near-wall turbulence structure}}.
  \jt{Phys. Fluids}  \bvol{8}~(1),  \pg{215--224}.

\bibitem[Port\'e-Agel {\em et~al.\/}(2001)Port\'e-Agel, Pahlow, Meneveau \&
  Parlange]{porte:pahlow:meneveau:par:2001}
{\sc \au{Port\'e-Agel, F.}, \au{Pahlow, M.}, \au{Meneveau, C.} \& \au{Parlange,
  M.~B.}} \yr{2001}  \at{{Atmospheric stability effect on subgrid-scale physics
  for large-eddy simulation}}.  \jt{Adv. Water Res.}  \bvol{24},
  \pg{1085--1102}.

\bibitem[Port{\'{e}}-Agel {\em et~al.\/}(2002)Port{\'{e}}-Agel, Parlange,
  Meneveau \& Eichinger]{porte:par:mene:eich:2001}
{\sc \au{Port{\'{e}}-Agel, F.}, \au{Parlange, M.~B.}, \au{Meneveau, C.} \&
  \au{Eichinger, W.~E.}} \yr{2002}  \at{{A priori field study of the
  subgrid-scale heat fluxes and dissipation in the atmospheric surface layer}}.
   \jt{J. Atmos. Sci.}  \bvol{58}~(18),  \pg{2673--2698}.

\bibitem[Richardson(1922)]{Richardson1922}
{\sc \au{Richardson, L.~F.}} \yr{1922} {\em Weather Prediction by Numerical
  Process\/}.  \publ{Cambridge University Press}.

\bibitem[Rogallo(1981)]{Rogallo81}
{\sc \au{Rogallo, R.~S.}} \yr{1981}  \bt{Numerical experiments in homogeneous
  turbulence}. Tech. Memo 81315.  \org{NASA}.

\bibitem[Schumann(1985)]{Schumann1985}
{\sc \au{Schumann, U.}} \yr{1985} Algorithms for direct numerical simulation of
  shear-periodic turbulence.  \bt{In {\em Ninth Int. Conf. on Numerical Methods
  in Fluid Dyn.\/} (ed. \ed{Soubbaramayer \& J.~P. Boujot})},  \st{Lecture
  Notes in Physics},  \vol{vol. 218},  \pg{pp. 492--496}.  \publ{Springer
  Berlin Heidelberg}.

\bibitem[Sekimoto {\em et~al.\/}(2016)Sekimoto, Dong \&
  Jim{\'{e}}nez]{sek:don:jim:15}
{\sc \au{Sekimoto, A.}, \au{Dong, S.} \& \au{Jim{\'{e}}nez, J.}} \yr{2016}
  \at{{Direct numerical simulation of statistically stationary and homogeneous
  shear turbulence and its relation to other shear flows}}.  \jt{Phys. Fluids}
  \bvol{28}~(3).

\bibitem[Sirovich \& Karlsson(1997)]{Sirovich1997}
{\sc \au{Sirovich, L.} \& \au{Karlsson, S.}} \yr{1997}  \at{Turbulent drag
  reduction by passive mechanisms}.  \jt{Nature}  \bvol{388},  \pg{753--755}.

\bibitem[Spalart {\em et~al.\/}(1991)Spalart, Moser \& Rogers]{Spalart1991}
{\sc \au{Spalart, P.~R.}, \au{Moser, R.~D.} \& \au{Rogers, M.~M.}} \yr{1991}
  \at{Spectral methods for the {N}avier-{S}tokes equations with one infinite
  and two periodic directions}.  \jt{J. Comput. Phys.}  \bvol{96},
  \pg{297--324}.

\bibitem[Vanderwel \& Tavoularis(2011)]{van:tav:11}
{\sc \au{Vanderwel, C.} \& \au{Tavoularis, S.}} \yr{2011}  \at{{Coherent
  structures in uniformly sheared turbulent flow}}.  \jt{J. Fluid Mech.}
  \bvol{689},  \pg{434--464}.

\bibitem[Veynante \& Vervisch(2002)]{Veynante2002}
{\sc \au{Veynante, D.} \& \au{Vervisch, L.}} \yr{2002}  \at{Turbulent
  combustion modeling}.  \jt{Prog. Energy Combust. Sci.}  \bvol{28}~(3),
  \pg{193 -- 266}.

\bibitem[Wu {\em et~al.\/}(2017)Wu, Moin, Wallace, Skarda, Lozano-Dur{\'a}n \&
  Hickey]{Wu2017}
{\sc \au{Wu, X.}, \au{Moin, P.}, \au{Wallace, J.~M.}, \au{Skarda, J.},
  \au{Lozano-Dur{\'a}n, A.} \& \au{Hickey, J.-P.}} \yr{2017}
  \at{Transitional{\textendash}turbulent spots and
  turbulent{\textendash}turbulent spots in boundary layers}.  \jt{Proc. Natl.
  Acad. Sci.}  \bvol{114}~(27),  \pg{E5292--E5299}.

\bibitem[Yang \& Lozano-Dur{\'a}n(2017)]{Yang2017}
{\sc \au{Yang, Xiyang I.~A.} \& \au{Lozano-Dur{\'a}n, Adri{\'a}n}} \yr{2017}
  \at{A multifractal model for the momentum transfer process in wall-bounded
  flows.}  \jt{J. Fluid Mech.}  \bvol{824}.

\bibitem[Young \& Read(2017)]{Young2017}
{\sc \au{Young, R. M.~B.} \& \au{Read, P.~L.}} \yr{2017}  \at{Forward and
  inverse kinetic energy cascades in {J}upiter's turbulent weather layer}.
  \jt{Nat. Phys.}  \bvol{13},  \pg{1135--1140}.

\end{thebibliography}
